\documentclass[structabstract]{aa}  
\usepackage{graphicx}
\usepackage{txfonts}
\usepackage{supertabular}
\usepackage{url}
\newcommand{\NH}{$N_{\rm H}$}
\newcommand{\NHUNIT}{H~cm$^{-2}$}

\newcommand{\hr}{$^{\rm h}$}
\newcommand{\mn}{$^{\rm m}$}
\newcommand{\Sec}{$^{\rm s}$}
\newcommand{\kalpha}{K$\alpha$~}
\newcommand{\xmm}{{\it XMM-Newton}~EPIC~}
\newcommand{\HII}{\mbox{H\,\textsc{ii}}}
\def\lsim{\lower.5ex\hbox{$\; \buildrel < \over \sim \;$}}
\def\gsim{\lower.5ex\hbox{$\; \buildrel > \over \sim \;$}}
\DeclareTextSymbol{\degre}{T1}{6}
\DeclareTextSymbol{\degre}{OT1}{23}
\begin{document}
\title{Nonthermal X-rays from low-energy cosmic rays: Application 
to the 6.4 keV line emission from the Arches cluster region}

\author{V. Tatischeff\inst{1}, A. Decourchelle\inst{2}, \and G. Maurin\inst{2,3}}

\offprints{V. Tatischeff}

\institute{Centre de Spectrom\'etrie Nucl\'eaire et de
Spectrom\'etrie de Masse, IN2P3/CNRS and Univ Paris-Sud, 91405 Orsay Campus, France
\\
              \email{Vincent.Tatischeff@csnsm.in2p3.fr}
         \and
             Service d'Astrophysique (SAp)/IRFU/DSM/CEA Saclay, Bt 709, 91191 Gif-sur-Yvette Cedex, France; 
	     Laboratoire AIM, CEA-IRFU/CNRS/Univ Paris Diderot, CEA Saclay, 91191 Gif sur Yvette, France
         \and
             Laboratoire d'Annecy le Vieux de Physique des Particules, Univ de Savoie, CNRS, BP~110, 
	     74941 Annecy-le-Vieux Cedex, France 
             }

\date{Received ...; accepted ...}

\abstract {The iron \kalpha line at 6.4~keV provides a valuable spectral diagnostic in several fields of X-ray 
astronomy. The line often results from the reprocessing of external hard X-rays by a neutral or low-ionized medium, 
but it can also be excited by impacts of low-energy cosmic rays.}
{This paper aims to provide signatures allowing identification of radiation from low-energy cosmic rays in 
X-ray spectra showing the 6.4 keV Fe \kalpha line.}
{We study in detail the production of nonthermal line and continuum X-rays by interaction of accelerated 
electrons and ions with a neutral ambient gas. Corresponding models are then applied to $XMM$-$Newton$ 
observations of the X-ray emission emanating from the Arches cluster region near the Galactic center.}
{Bright 6.4~keV Fe line structures are observed around the Arches cluster. This emission is very likely 
produced by cosmic rays. We find that it can result from the bombardment of molecular gas by energetic ions, but 
probably not by accelerated electrons. Using a model of X-ray production by cosmic-ray ions, 
we obtain a best-fit metallicity of the ambient medium of $1.7 \pm 0.2$ times the solar metallicity. A large 
flux of low-energy cosmic ray ions could be produced in the ongoing supersonic collision between the star cluster 
and an adjacent molecular cloud. We find that a particle acceleration efficiency in the resulting shock system of 
a few percent would give enough power in the cosmic rays to explain the luminosity of the nonthermal X-ray 
emission. Depending on the unknown shape of the kinetic energy distribution of the fast ions above
$\sim$1~GeV~nucleon$^{-1}$, the Arches cluster region may be a source of high-energy $\gamma$-rays detectable with 
the $Fermi$ Gamma-ray Space Telescope.}
{At present, the X-ray emission prominent in the 6.4~keV Fe line emanating from the Arches cluster region probably offers 
the best available signature for a source of low-energy hadronic cosmic rays in the Galaxy. }

\titlerunning{Nonthermal X-rays from low-energy cosmic rays}

   \maketitle
%
%________________________________________________________________

\section{Introduction}

The Fe \kalpha line at 6.4~keV from neutral to low-ionized Fe atoms is an important 
probe of high-energy phenomena in various astrophysical sites. It is produced by 
removing a K-shell electron, either by hard X-ray photoionization or by the
collisional ionization induced by accelerated particles, rapidly followed by an 
electronic transition from the L shell to fill the vacancy. The Fe \kalpha line 
emitted from a hot, thermally-ionized plasma at ionization equilibrium is generally 
in the range 6.6--6.7~keV depending on the plasma temperature. 

The 6.4 keV Fe \kalpha line is a ubiquitous emission feature in the X-ray spectra of active 
galactic nuclei (Fukazawa et al. \cite{fuk11}). It is also commonly detected from 
high-mass X-ray binaries (Torrej{\'o}n et al. \cite{tor10}) and some cataclysmic 
variables (Hellier \& Mukai \cite{hel04}). In these objects, the line is attributed 
to the fluorescence from photoionized matter in the vicinity of a compact, bright X-ray 
source (George \& Fabian \cite{geo91} and references therein). The 6.4~keV line 
is also detected in solar flares (Culhane et al. \cite{cul81}), low-mass, flaring 
stars (Osten et al. \cite{ost10}), massive stars ($\eta$~Car; Hamaguchi et al. 
\cite{ham07}), young stellar objects (Tsujimoto et al. \cite{tsu05}), supernova 
remnants (RCW~86; Vink et al. \cite{vin97}), and molecular clouds in the Galactic 
center region (Ponti et al. \cite{pon10}). 

One of the best studied cases of this emission is the Sun. The observed line intensity 
and light curve during several flares suggest that excitation of Fe atoms occurs 
mainly by photoionization induced by flare X-rays with, however, an 
additional contribution in some impulsive events from collisional ionization by 
accelerated electrons (Zarro et al. \cite{zar92}). A contribution from 
collisional ionization by accelerated electrons is also discussed for the emission 
at 6.4~keV from low-mass flaring stars (Osten et al. \cite{ost10}) and young stellar 
objects (Giardino et al. \cite{gia07}). 

The 6.4~keV line emission from the Galactic center (GC) region was predicted by Sunyaev et 
al. (\cite{sun93}) before being discovered by Koyama et al. (\cite{koy96}). These authors 
suggest that the neutral Fe \kalpha line can be produced in molecular clouds, 
together with nonthermal X-ray continuum radiation, as a result of reprocessed emission 
of a powerful X-ray flare from the supermassive black hole Sgr~A$^\ast$. Recent 
observations of a temporal variation in the line emission from various clouds of the 
central molecular zone can indeed be explained by a long-duration flaring activity 
of Sgr~A$^\ast$ that ended about 100 years ago (Muno et al. \cite{mun07}; Inui et al. 
\cite{inu09}; Ponti et al. \cite{pon10}; Terrier et al. \cite{ter10}). Some data also 
suggest there is a background, stationary emission in the Fe line at 6.4 keV 
(Ponti et al. \cite{pon10}), which might be due to the interaction of cosmic rays with 
molecular clouds. Observations showing a spatial correlation between the X-ray line emission 
and nonthermal radio filaments have been interpreted as evidence of a large population of 
accelerated electrons in the GC region (Yusef-Zadeh et al. \cite{yus02a,yus07}). Alternatively, 
Dogiel et al. (\cite{dog09,dog11}) suggest that the neutral or low ionization Fe \kalpha 
line from this region could be partly excited by subrelativistic protons generated 
by star accretion onto the central supermassive black hole.

Low-energy cosmic-ray electrons propagating in the interstellar medium (ISM) have also been 
invoked to explain the presence of a nonthermal continuum and a weak line at 6.4~keV in the 
spectrum of the Galactic ridge X-ray background (Valina et al. \cite{val00}). But most of  
Galactic ridge X-ray emission has now been resolved into discrete sources, probably cataclysmic 
variables and coronally active stars (Revnivtsev et al. \cite{rev09}). Therefore, it is likely 
that the 6.4~keV line in the Galactic ridge spectrum is produced in these sources and not in the 
ISM. 

In this paper, we study in detail the production of nonthermal line and continuum X-rays 
by interaction of accelerated electrons, protons, and $\alpha$-particles with a neutral 
ambient gas. Our first aim is to search for spectral signatures that allow
identification of cosmic-ray-induced X-ray emission. We then apply the developed models 
to the X-ray emission from the Arches cluster region near the GC. 

The Arches cluster is an extraordinary massive and dense cluster of young stars, with 
possibly 160 O-type stars with initial masses greater than 20~$M_\odot$ and an average 
mass density of $\sim$$3\times 10^5~M_\odot$~pc$^{-3}$ (Figer et al. \cite{fig02}). The
X-ray emission from the cluster is a mix of thermal and nonthermal radiations. The 
thermal emission is thought to arise from multiple collisions between strong winds 
from massive stars (Yusef-Zadeh et al. \cite{yus02b}; Wang et al. \cite{wan06}). This
interpretation was recently reinforced by the detection with the {\it XMM-Newton} 
observatory of X-ray flaring activity within the cluster, which likely originates in 
one or more extreme colliding wind massive star binaries (Capelli et al. \cite{cap11a}). 
Diffuse nonthermal emission prominent in the Fe K$\alpha$ 6.4-keV line has also been detected 
from a broad region around the cluster (Wang et al. \cite{wan06}; Tsujimoto et al. 
\cite{tsu07}; Capelli et al. \cite{cap11b}). Wang et al. (\cite{wan06}) suggest from a 
100-ks {\it Chandra} observation that this component may be produced by interaction of 
low-energy cosmic-ray electrons with a dense gas in a bow shock resulting from the 
supersonic collision of the star cluster with a molecular cloud. In this 
scenario, the nonthermal electrons may be accelerated in the bow-shock system itself  
and/or in shocked stellar winds within the Arches cluster. The latter assumption is 
supported by the detection with the Very Large Array (VLA) of diffuse nonthermal radio 
continuum emission from the cluster (Yusef-Zadeh et al. \cite{yus03}). However, Tsujimoto 
et al. (\cite{tsu07}) show from {\it Suzaku} observations using preliminary 
calculations of Tatischeff (\cite{tat03}) that this scenario would require a very high 
Fe abundance in the ambient medium, about four to five times the solar value. Capelli et al. 
(\cite{cap11b}) have recently favored a photoionization origin for the 6.4 keV line 
from the Arches cluster region, although not excluding a production by low-energy 
cosmic-ray electrons and/or protons. We show in the present work that the 6.4 keV 
line from this region is indeed most likely excited by subrelativistic ion collisions. 

The plan of the paper is as follows. In Sect.~2, we theoretically study the production 
of nonthermal line and continuum X-rays by interaction of accelerated electrons and ions 
with a neutral ambient gas. In Sect.~3, we present the {\it XMM-Newton} observations of the 
Arches cluster region and describe the data reduction technique we employed. In Sect.~4, we 
study the temporal variability of the 6.4 keV line detected from a broad region surrounding 
the star cluster. In Sect.~5, we present a detailed spectral analysis of the {\it XMM-Newton} 
data that uses the newly developed cosmic-ray models. The origin of the detected thermal 
and nonthermal radiations is discussed in Sect.~6, where we argue that the 6.4 keV line emission 
in the vicinity of the star cluster is produced by a large population of low-energy cosmic ray 
ions. The acceleration source of these particles is discussed in Sect. 7. In Sect.~8, 
we estimate the ionization rate induced by the fast ions in the ambient medium. In Sect.~9,
we investigate the gamma-ray emission from this region. A summary is finally given in Sect.~10.

\section{Nonthermal X-rays from low-energy cosmic rays}

The X-ray production is calculated in the framework of a generic, steady-state, slab 
model, in which low-energy cosmic rays (LECRs) penetrate a cloud of neutral gas at a 
constant rate. The fast particles slow down by ionization and radiative energy losses in the cloud and can 
either stop or escape from it depending on their path length in the ambient medium, $\Lambda$, which 
is a free parameter of the model. There are three other free parameters that can be studied from 
spectral fitting of X-ray data: the minimum energy of the CRs entering the 
cloud, $E_{\rm min}$, the power-law index of the CR source energy spectrum, $s$, and the metallicity 
of the X-ray emission region, $Z$. More details about the cosmic-ray interaction model are given in 
Appendix~A. 

In Appendices~B and C, we describe the atomic processes leading to X-ray continuum and line production 
as a result of accelerated electron and ion impacts. At this stage, we neglect the broad 
lines that can arise from atomic transitions in fast C and heavier ions following electron captures 
and excitations (Tatischeff et al. \cite{tat98}). We only study the production of the narrower lines 
that result from K-shell vacancy production in the ambient atoms. We consider the K$\alpha$ and 
K$\beta$ lines from ambient C, N, O, Ne, Mg, Si, S, Ar, Ca, Fe, and Ni. We now examine the properties 
of the most important of these narrow lines in detail, the one at 6.4 keV from ambient Fe. 

\subsection{LECR electrons}

   \begin{figure}
   \centering
   \includegraphics[width=7.5cm]{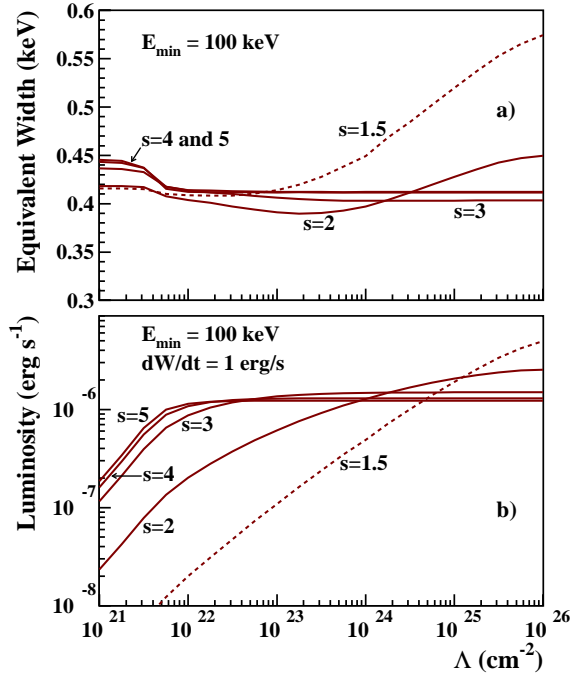}
      \caption{Calculated {\bf a)} EW and {\bf b)} luminosity of the 6.4 keV 
      Fe \kalpha line produced by LECR electrons as a function of the path length of 
      the primary electrons injected in the X-ray production region, for five values of the 
      electron spectral index $s$ (Eq.~(\ref{eq7})). The ambient medium is assumed to have a
      solar composition and the electron minimum energy $E_{\rm min}=100$~keV. In panel~b, 
      the luminosity calculations are normalized to a total power of 1~erg~s$^{-1}$ 
      injected by the fast primary electrons in the ambient medium. 
              }
         \label{Fig4}
   \end{figure}

   \begin{figure}
   \centering
   \includegraphics[width=7.5cm]{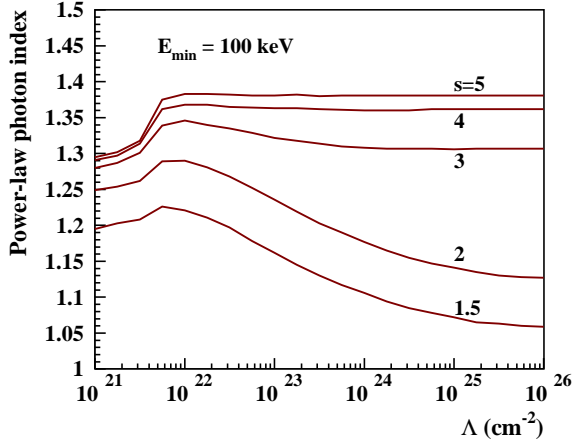}
      \caption{Slope at 6.4 keV of the bremsstrahlung continuum emission 
      produced by LECR electrons, as a function of the path length of the primary
      electrons injected in the X-ray production region, for five values of the  
      spectral index $s$. The electron minimum energy is taken to be $E_{\rm min}=100$~keV.
      	}
         \label{Fig5}
   \end{figure}

   \begin{figure}
   \centering
   \includegraphics[width=7.5cm]{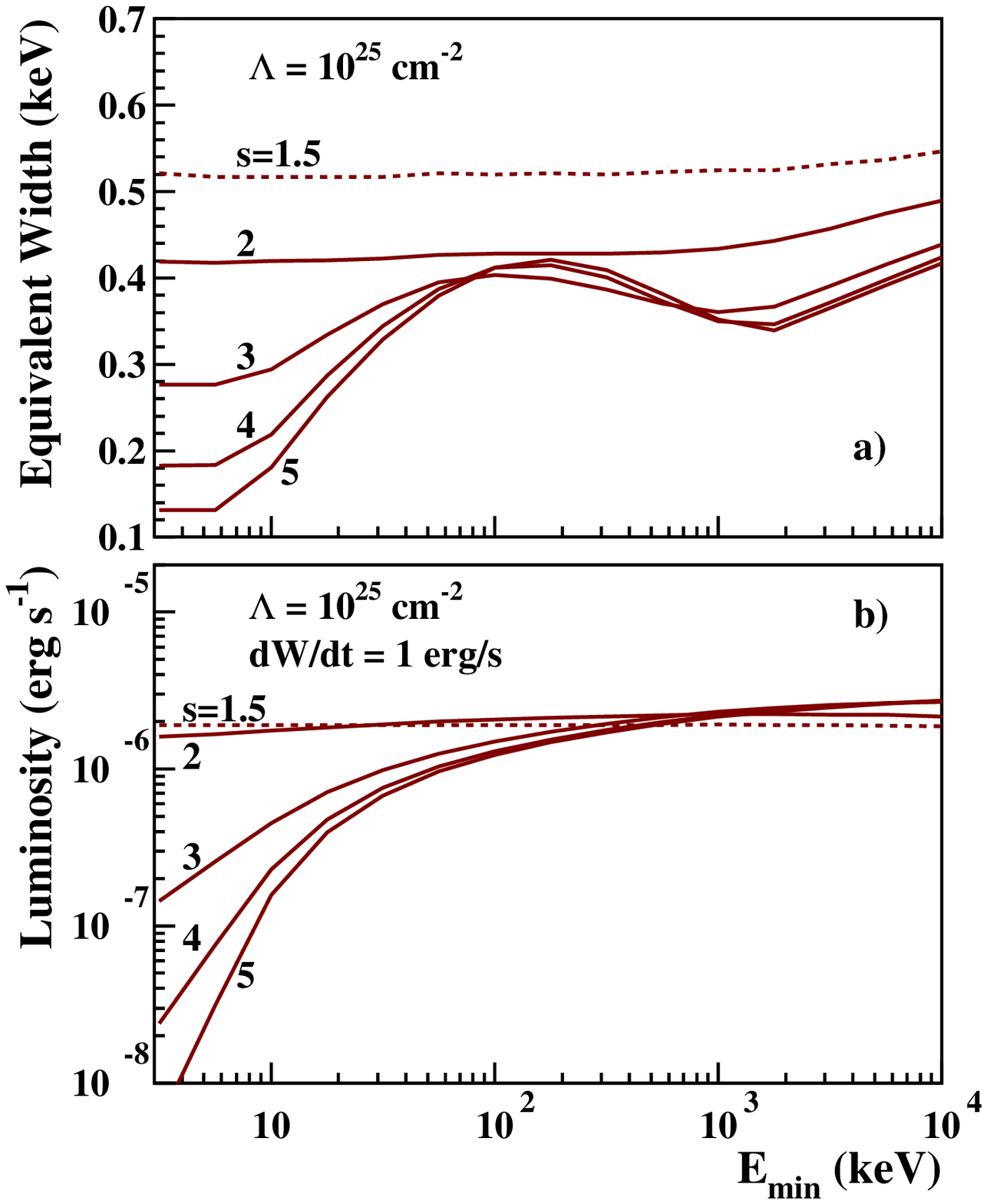}
      \caption{Same as Fig.~1 but as a function of the electron minimum energy, 
      for $\Lambda=10^{25}$~cm$^{-2}$.
              }
         \label{Fig6}
   \end{figure}

   \begin{figure}
   \centering
   \includegraphics[width=7.5cm]{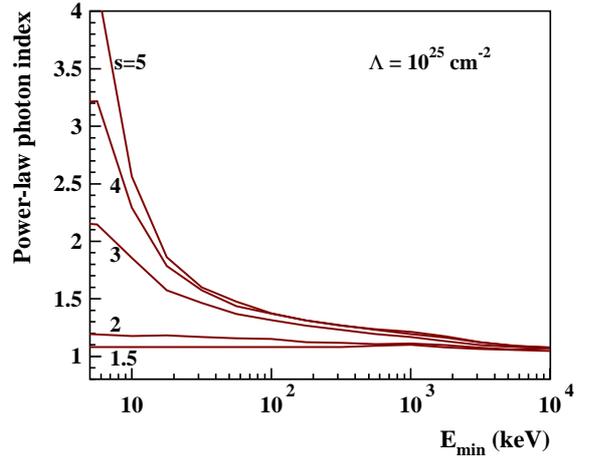}
      \caption{Same as Fig.~2 but as a function of the electron minimum energy, 
      for $\Lambda=10^{25}$~cm$^{-2}$. 
              }
         \label{Fig7}
   \end{figure}

We present in Figs.~\ref{Fig4}--\ref{Fig7} characteristic properties of the X-ray spectrum 
resulting from LECR electron interactions around the neutral Fe \kalpha line. All the calculations 
were performed for an ambient medium of solar composition (i.e. $Z=Z_\odot$ where $Z_\odot$ is the 
solar metallicity). Figures~\ref{Fig4} and \ref{Fig6} show the equivalent width (EW) and luminosity 
of the 6.4 keV line, whereas Figs.~\ref{Fig5} and \ref{Fig7} show the slope of the underlying continuum 
emission at the same energy. The former two quantities depend linearly on the metallicity, whereas the 
continuum emission, which is produced by electron bremsstrahlung in ambient H and He, is independent of $Z$. 

We see in Figs.~\ref{Fig4}a and \ref{Fig6}a that the EW of the Fe \kalpha line is generally
lower than $\sim$0.45$\times (Z/Z_\odot)$~keV. The only exception is for $s = 1.5$ and 
$\Lambda > 10^{24}$~cm$^{-2}$. But in all cases, we expect the EW to be lower than $0.6\times (Z/Z_\odot)$~keV. 
This result constitutes a strong constraint for a possible contribution of LECR electrons to the 
6.4~keV line emission from the GC region, because the observed  EW is $>1$~keV in some places (see Sect.~5) 
and sometimes equal to $\sim$2~keV (see, e.g., Revnivtsev et al. \cite{rev04}). 

The results shown in Figs.~\ref{Fig4}a and \ref{Fig6}a were obtained without considering 
the additional fluorescent line emission that can result from photoionization 
of ambient Fe atoms by bremsstrahlung X-rays $>$7.1~keV emitted in the cloud. This 
contribution can be estimated from the Monte-Carlo simulations of Leahy \& Creighton 
(\cite{lea93}), who studied the X-ray spectra produced by reprocessing of a 
power-law photon source surrounded by cold matter in spherical geometry. For 
the power-law photon index $\alpha=1$, the simulated EW of the neutral Fe \kalpha 
line can be satisfactorily approximated by
\begin{equation} 
{\rm EW_{LC93}} \approx 0.07\times (Z/Z_\odot) \times
\bigg({N_{\rm H}^C \over 10^{23}~{\rm cm}^{-2}} \bigg)~~{\rm keV}~,
\label{eq12}
\end{equation}
as long as the radial column density of the absorbing cloud, $N_{\rm H}^C$, is lower 
than 10$^{24}$~cm$^{-2}$. For $\alpha=2$, we have ${\rm EW_{LC93}} \approx 0.03\times (Z/Z_\odot) 
\times (N_{\rm H}^C / 10^{23}~{\rm cm}^{-2})$~keV. Thus, we see by comparing these 
results with those shown in Figs.~\ref{Fig4}a and \ref{Fig6}a that the additional 
contribution from internal fluorescence is not strong for 
$N_{\rm H}^C \lsim 5 \times 10^{23}$~cm$^{-2}$. 

   \begin{figure*}
   \centering
   \includegraphics[width=0.7\textwidth]{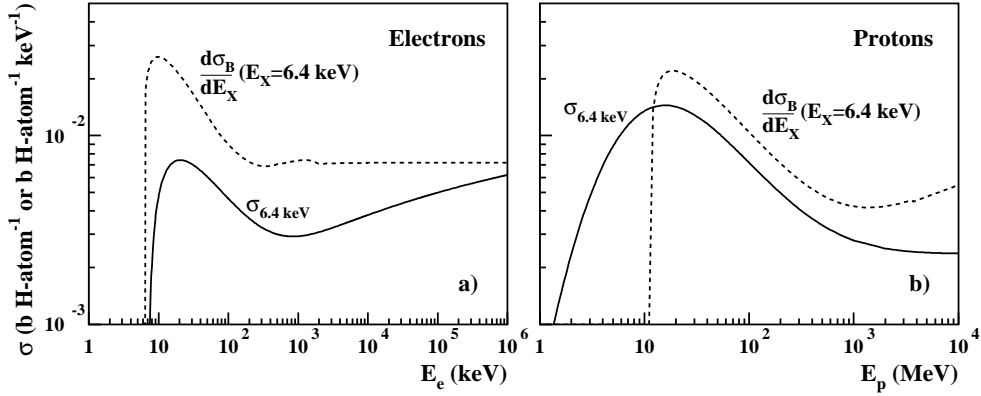}
      \caption{Cross sections involved in the calculation of the Fe \kalpha line EW. 
      {\it Solid lines}: cross sections (in units of barn per ambient H-atom) 
      for producing the 6.4 keV line by the impact of fast electrons ({\it left panel}) 
      and protons ({\it right panel}), assuming solar metallicity. {\it Dashed lines}: 
      differential cross section (in barn per H-atom per keV) for producing 6.4 keV 
      X-rays by bremsstrahlung of fast electrons ({\it left panel}) and inverse 
      bremsstrahlung from fast protons ({\it right panel}), in a medium composed of 
      H and He with H/He=0.1. The ratio of these two cross sections gives the 
      EW of the 6.4 keV line (in keV) for a mono-energetic beam of 
      accelerated particles. 
              }
         \label{Fig8}
   \end{figure*}

As shown in Figs.~\ref{Fig4}b and \ref{Fig6}b, the production of 6.4~keV line photons by 
LECR electron interactions is relatively inefficient: the radiation yield 
$R_{6.4~{\rm keV}}=L_X(6.4~{\rm keV}) / (dW_e/dt)$ is always lower than $3 \times 10^{-6} \, (Z/Z_\odot)$, 
which implies that a high kinetic power in CR electrons should generally be needed to produce
an observable K$\alpha$ line from neutral or low-ionized Fe atoms. For example, the total luminosity 
of the 6.4~keV line emission from the inner couple of hundred parsecs of our Galaxy is 
$>6 \times 10^{34}$~erg~s$^{-1}$ (Yusez-Zadeh et al. \cite{yus07}), such that 
$dW_e/dt > 2 \times 10^{40}$~erg~s$^{-1}$ would be needed if this emission was 
entirely due to LECR electrons (assuming the ambient medium to be of solar metallicity). 
Such power would be comparable to that contained in CR protons in the entire Galaxy. 

On the other hand, Fig.~\ref{Fig4}b shows that LECR electrons can produce a
significant Fe K$\alpha$ line (i.e. $R_{6.4~{\rm keV}} \sim 10^{-6}$) in diffuse 
molecular clouds with $N_{\rm H}^C < 10^{22}$~cm$^{-2}$, especially in the 
case of strong particle diffusion for which $\Lambda$ can be much larger than 
$N_{\rm H}^C$ (see Appendix ~A). An observation of a 6.4~keV line emission from a cloud 
with $N_{\rm H}^C \sim 10^{21}$~cm$^{-2}$ would potentially be a promising 
signature of LECR electrons, since the efficiency of production of this line by 
hard X-ray irradiation of the cloud would be low: the ratio of the 6.4~keV line 
flux to the integrated flux in the incident X-ray continuum above 7.1~keV (the 
K-edge of neutral Fe) is only $\sim 10^{-4}$ for $N_{\rm H}^C = 10^{21}$~cm$^{-2}$
(Yaqoob et al. \cite{yaq10}). 

Figures~\ref{Fig5} and \ref{Fig7} show the slope $\Gamma$ of the bremsstrahlung 
continuum emission, as obtained from the derivative of the differential X-ray production rate
$\partial(dQ_X/dt)/\partial E_X$
taken at 6.4~keV. We see that, for $E_{\rm min}>100$~keV, $\Gamma$ is lower than 1.4 
regardless of $s$ and $\Lambda$. This is because bremsstrahlung X-rays $< 10$~keV are 
mainly produced by LECR electrons $< 100$~keV, and the equilibrium spectrum of these 
electrons is hard (see Fig.~\ref{Fig2}) and depends only weakly on the distribution 
of electrons injected in the ambient medium at higher energies. 

Thus, after having studied the influence of the free parameters over broad ranges, 
we can summarize the main characteristics of the X-ray emission produced by LECR 
electrons as follows. First, the continuum radiation should generally be hard, 
$\Gamma < 1.4$, provided that nonthermal electrons $\lsim 100$~keV are not able to 
escape from their acceleration region and penetrate denser clouds. Secondly, 
the EW of the 6.4 keV Fe \kalpha line is predicted to be
$\sim (0.3~$--$~0.5) \times (Z/Z_\odot)$~keV, whatever the electron acceleration spectrum 
and transport in the ambient medium.

The reason that the EW of the 6.4~keV line is largely independent of the electron 
energy distribution is given in Fig.~\ref{Fig8}a, which shows the relevant cross 
sections for this issue. The solid line is the cross section for producing 
6.4 keV Fe K$\alpha$ X-rays expressed in barn per ambient H-atom, that is 
$a_{\rm Fe} \times \sigma_{e\rm Fe}^{K\alpha}$ (see Eq.~(\ref{eq11})). The dashed 
line is the differential cross section for producing X-rays of the same energy by electron
bremsstrahlung. The Fe line EW produced by a given electron energy distribution is 
obtained from the ratio of the former cross section to the latter, convolved over 
that distribution. We see that the two cross sections have similar shapes, in 
particular similar energy thresholds, which explains why the EW of the 6.4~keV line 
depends only weakly on the electron energy. However, the cross section for 
producing the 6.4~keV line increases above 1~MeV as a result of relativistic effects 
in the K-shell ionization process (Quarles~\cite{qua76}; Kim et al.~\cite{kim00b}), 
which explains why the EW slightly increases with the hardness of the electron 
source spectrum (see Fig.~\ref{Fig6}a). 

Figure~\ref{Fig8}b shows the same cross sections but for proton impact. The 
calculation of these cross sections are presented in Appendix C. 
We see that the cross section for the line production has a lower energy threshold 
than that for the bremsstrahlung continuum. We thus expect that LECR protons with 
a relatively soft source spectrum can produce a higher EW of the 6.4~keV line than
the electrons (see also Dogiel et al. \cite{dog11}). 

\subsection{LECR ions}

   \begin{figure}
   \centering
   \includegraphics[width=7.5cm]{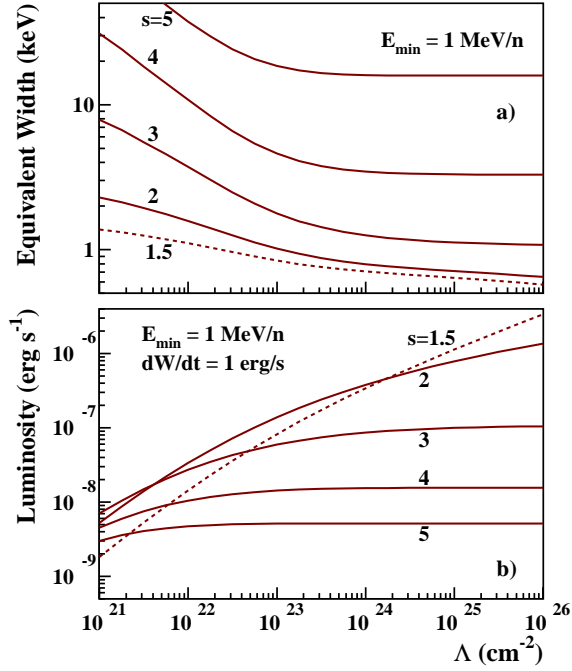}
      \caption{Calculated {\bf a)} EW and {\bf b)} luminosity of the 6.4 keV 
      Fe \kalpha line produced by LECR ions as a function of the path length of the CRs 
      injected in the X-ray production region, for five values of the CR source spectral index 
      $s$ (Eq.~(\ref{eq7})). The ambient medium is assumed to have a solar composition, and
      the minimum energy of the CRs that penetrate this medium is $E_{\rm min}=1$~MeV~nucleon$^{-1}$. 
      In panel~b, the luminosity calculations are normalized to a total power of 1~erg~s$^{-1}$ 
      injected by the fast primary protons in the ambient medium. 
	}
         \label{Fig11}
   \end{figure}

   \begin{figure}
   \centering
   \includegraphics[width=7.5cm]{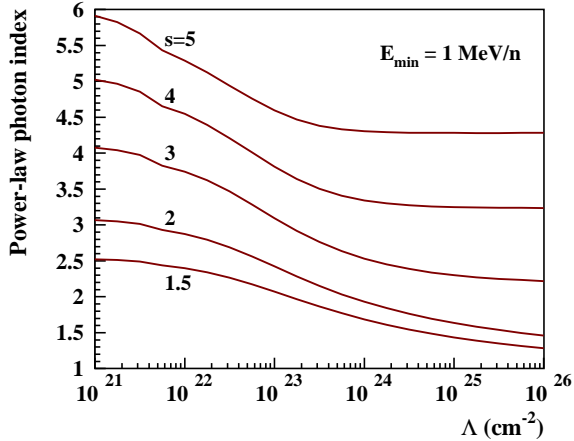}
      \caption{Slope at 6.4 keV of the continuum emission produced by 
      LECR ions as a function of the path length of the fast ions in the X-ray 
      production region, for five values of the spectral index $s$. The minimum 
      energy of injection is taken to be $E_{\rm min}=1$~MeV~nucleon$^{-1}$.
              }
         \label{Fig12}
   \end{figure}

   \begin{figure}
   \centering
   \includegraphics[width=7.5cm]{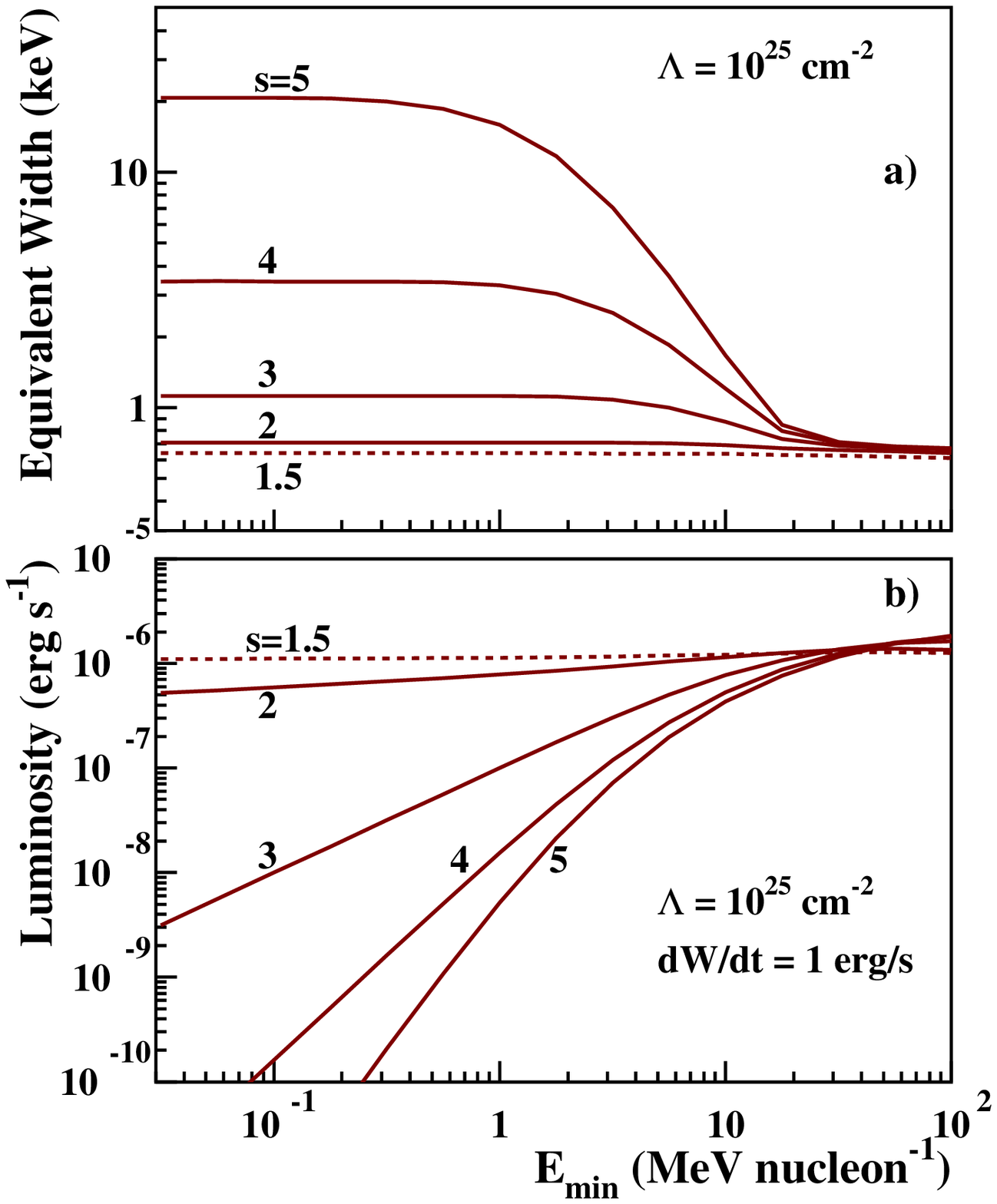}
      \caption{Same as Fig.~6 but as a function of the minimum energy of injection, 
      for $\Lambda=10^{25}$~cm$^{-2}$.
              }
         \label{Fig13}
   \end{figure}

   \begin{figure}
   \centering
   \includegraphics[width=7.5cm]{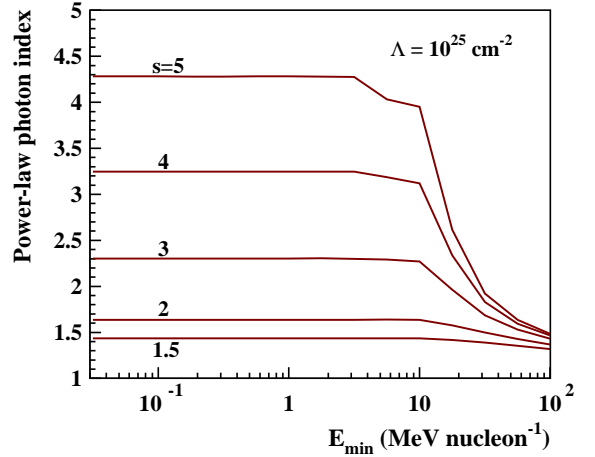}
      \caption{Same as Fig.~7 but as a function of the minimum energy of injection, 
      for $\Lambda=10^{25}$~cm$^{-2}$. 
              }
         \label{Fig14}
   \end{figure}

Figures~\ref{Fig11}--\ref{Fig14} present characteristic properties of the X-ray spectrum 
resulting from LECR ion interactions around the neutral Fe \kalpha line. As before, all the 
calculations were done for an ambient medium of solar composition. The most remarkable result is 
that fast ions with a soft source spectrum can produce very high EW of the 6.4~keV line 
(Figs.~\ref{Fig11}a and \ref{Fig13}a). However, we see in Fig.~\ref{Fig13}a that the line EW 
becomes almost independent of $s$ for $E_{\rm min} \gsim 20$~MeV~nucleon$^{-1}$. This is because 
(i) the cross section for producing continuum X-rays at 6.4~keV and the one for the neutral Fe \kalpha 
line have similar shapes above 20~MeV~nucleon$^{-1}$ (see Fig.~\ref{Fig8}b) and (ii) the CR equilibrium 
spectrum below $E_{\rm min}$ only weakly depends on $s$ and $\Lambda$. 

As shown in Fig.~\ref{Fig11}b, for $E_{\rm min} = 1$~MeV~nucleon$^{-1}$, the radiation yield 
$R_{6.4~{\rm keV}}$ can reach $\sim$$10^{-6}$ only for relatively hard source spectra with 
$s \leq 2$ and for $\Lambda \gsim 10^{25}$~cm$^{-2}$, which should generally mean 
strong particle diffusion in the X-ray production region. For such a high CR path length, 
the X-rays are mainly produced in thick-target interactions. An efficiency of $\sim$$10^{-6}$ 
in the production of the 6.4~keV line can also be achieved with a softer CR source spectrum, 
if $E_{\rm min}$ is in the range 10--100~MeV~nucleon$^{-1}$ (Fig.~\ref{Fig13}b). This is because
then most of the CRs are injected into the X-ray production region at energies where the 
cross section for producing Fe K$\alpha$ X-rays is highest (see Fig.~\ref{Fig8}b). But in any 
case, we find that to get $R_{6.4~{\rm keV}} \sim 10^{-6}$ it requires 
$\Lambda \gsim 10^{24}$~cm$^{-2}$ (for solar metallicity). It is another difference from the 
production of nonthermal X-rays by LECR electrons, for which $R_{6.4~{\rm keV}}$ can reach 
$\sim$$10^{-6}$ for $\Lambda$ as low as 10$^{22}$~cm$^{-2}$ (Fig.~\ref{Fig4}b). 

Figures~\ref{Fig12} and \ref{Fig14} show that the characteristic power-law slope of the continuum 
emission around 6.4~keV can vary from $\sim$1 to $\sim$6 for $s$ in the range 1.5--5. However, for
$s \leq 2$, which is expected for strong shock acceleration of nonrelativistic particles, and 
$\Lambda > 10^{24}$~cm$^{-2}$, which can result from strong particle diffusion in the cloud, we expect 
$\Gamma$ between 1.3 and 2. For these conditions, the EW of the neutral Fe \kalpha line is 
predicted to be in the narrow range $(0.6~$--$~0.8) \times (Z/Z_\odot)$~keV (Figs.~\ref{Fig11}a and 
\ref{Fig13}a). This result could account for EW $\gsim 1$~keV as observed from regions near the 
GC, provided that the diffuse gas there has a super-solar metallicity 
(i.e. $Z > Z_\odot$). 

\section{XMM-Newton observations and data reduction}

\subsection{Data reduction and particle background substraction}

\begin{table*}[htbp]
\begin{center}
\caption{Summary of the XMM-Newton/EPIC observations available for the Arches cluster.}
\label{Table1}
\begin{tabular}{|c|c|r|r|c|c|c|c|c|c|c|c|c|}
\hline
       &  & \multicolumn{4}{c|}{Good time exposure} &   &  \multicolumn{6}{c|}{coverage (\%)}   \\
\cline{8-13}
Date       & Observation & \multicolumn{4}{c|}{(ks)} & MOS  &  \multicolumn{3}{c|}{Cloud} & \multicolumn{3}{c|}{Cluster} \\
\cline{3-6}
\cline{8-13}
                   & ID                                & M1 & M2  & \multicolumn{2}{|c|}{pn}  & Noisy CCD &  M1 & M2 &  pn & M1 & M2 & pn \\
\hline
\hline
2000-09-19 & 0112970401 &  21 & 21  &  \multicolumn{2}{|c|}{16}         & -                   &  100  & 100   &   86 & 100  & 100   &100 \\   
2000-09-21 & 0112970501 &    7 &   9  &  \multicolumn{2}{|c|}{3}           & -                   &   96   & 99   &   87 & 99  & 100   &100 \\   
2001-09-04 & 0112972101 &  20 & 20  &  \multicolumn{2}{|c|}{17}         & -                   &   100 & 100   &   40 & 100  & 100   &   21 \\ 
2002-02-26 & 0111350101 &  43 & 42  &  \multicolumn{2}{|c|}{36}         & -                   &   100 & 100   &   56 & 100  & 100   &   94 \\  
2002-10-02 & 0111350301 &    6 &   6  &  \multicolumn{2}{|c|}{3}           & -                   &   99 & 100   &   47 & 100  & 100   &   82 \\ 
2004-03-28 & 0202670501 &  10 &  14 &  \multicolumn{2}{|c|}{2}           & -                   &   100 & 100   & 100 & 100  & 100   &100 \\   
2004-03-30 & 0202670601 &  27 &  27 &  \multicolumn{2}{|c|}{17}         & -                   &   100 & 100   &   99 & 100  & 100   &100 \\   
2004-08-31 & 0202670701 &  64 &  74 &  \multicolumn{2}{|c|}{40}         & M2-5            &   100 & 100   & 100 & 100  & 100   &100 \\   
2004-09-02 & 0202670801 &  83 &  89 &  \multicolumn{2}{|c|}{50}         & M2-5            &   100 & 100   & 100 & 100  & 100   &100 \\   
2006-02-27 & 0302882601 &   2  &    2 &  \multicolumn{2}{|c|}{1}           & -                   &   0     & 100   &   55 & 0  & 100   &  91 \\
2006-09-08 & 0302884001 &    6 &    6 &  \multicolumn{2}{|c|}{4}           & M1-4            &   100 & 99   &   33 & 100  & 100   &   7 \\ 
2007-02-27 & 0506291201 &  18 &  20 &  \multicolumn{2}{|c|}{\tablefootmark{a}} & -                   &   100 & 87     &  \tablefootmark{a}    & 100  & 96   &  \tablefootmark{a} \\     
2007-03-30 & 0402430701 &  23 &  24 &  \multicolumn{2}{|c|}{16}         & M2-5            &   0     & 100   & 100 & 0  & 100   &100 \\ 
2007-04-01 & 0402430301 &  54 &  58 &  \multicolumn{2}{|c|}{29}         & M1-4, M2-5 &   0     & 100    &   99 & 0  & 100   &100 \\ 
2007-04-03 & 0402430401 &  39 &  40 &  \multicolumn{2}{|c|}{23}         & M1-4            &   0     & 100   & 100 & 0  & 100   &100 \\ 
2007-09-06 & 0504940201 &    8 &  9   &  \multicolumn{2}{|c|}{5}           & M2-5            &   98   & 99   &   34 & 98  & 100   &  23 \\ 
2008-03-04 & 0511000301 &    4 &  4   &  \multicolumn{2}{|c|}{2}           & -                   &   0     & 100  &    64 & 0  & 100   &100 \\
2008-03-23 & 0505670101 &  68 &  73 &  \multicolumn{2}{|c|}{51}         & M1-4, M2-5  &   0     & 100  & \tablefootmark{b} & 0  & 100   & \tablefootmark{b} \\ 
2008-09-23 & 0511000401 &    4 &    4 &  \multicolumn{2}{|c|}{4}           & M1-4, M2-5  &   99   & 99  &    41 & 98  & 100   &  41 \\ 
2009-04-01 & 0554750401 & 32  &  32 &  \multicolumn{2}{|c|}{24}         & -                   &    0    & 100  &    98 & 0  & 100   &100 \\ 
2009-04-03 & 0554750501 & 40  &  41 &  \multicolumn{2}{|c|}{31}         & M1-4            &    0    & 100  &    99 & 0  & 100   &100 \\ 
2009-04-05 & 0554750601 & 36  &  37 &  \multicolumn{2}{|c|}{25}         & M1-4            &    0    & 100  &    99 & 0  & 100   &100 \\ 
\hline
\cline{1-6}
Total            & Imaging                    & 615 & 652 & \multicolumn{2}{|c|}{399} & \multicolumn{7}{c}{} \\
\cline{2-6}
exposure (ks)  &                             &         &       & Cloud    &  Cluster             & \multicolumn{7}{c}{}  \\
\cline{5-6}
                      & Spectroscopy        & 317  & 652 & 276   &  315                      & \multicolumn{7}{c}{} \\
\cline{1-6}
\end{tabular}
\end{center}
\tablefoot{
For each observation and instrument, we report the total good-time interval exposure after flare screening. For a number of observations, the Arches cluster lies at the position of the CCD 6 of MOS 1, which is out of service. We indicate the CCDs identified as noisy for MOS (data not used for the analysis) and the spatial coverage of the pn (incomplete due to dead columns)  for our two spectral extraction regions (see Fig.~\ref{Fig15}). Only observations with a spatial coverage of the region greater than 85\% were used for the spectral analysis.\\
\tablefoottext{a}{The pn camera was in timing mode for this observation.}
\tablefoottext{b}{The pn data of this observation were not taken into account for the spectrum extraction, because the pn spatial coverage of the background region (Fig.~\ref{Fig15}) was to low.}
}
\end{table*}

For our analysis, we have considered all public \xmm observations encompassing the Arches cluster (R.A.= 17\hr45\mn50\Sec, Dec=-28\degre49'20"). The criteria were to have more than 1.5 ks observation time available for each camera, to be in full frame or extended full frame mode, and to use the medium filter. The data was reduced using the SAS software package, version 10.0. Calibrated-event files were produced using the tasks {\small EMCHAIN} for the MOS cameras and {\small EPPROC} for the pn camera. We excluded from the analysis the period contaminated by soft proton flares, by using an automatic {\it $3\sigma$-clipping} method (\cite{Pratt}). Table~\ref{Table1} provides the list of the 22 selected observations and their respective observing time per instrument after flare rejection.

We searched for any anomalous state of MOS CCD chips (\cite{Kuntz}) by performing a systematic inspection of the images and spectra of each chip in the 0.3--1~keV energy band. We identified 14 occurrences of a noisy chip in the list of observations (see Table~\ref{Table1}). The chips affected in our observations by a high-level, low-energy background state are CCD 4 of MOS 1 and CCD 5 of MOS 2. We excluded data of those chips from our analysis when they were noisy.

For MOS cameras, we selected events with {\small PATTERN}~$ \leq 12$. Only events with {\small PATTERN}~$ \leq 4$ were kept 
for the pn instrument. Depending on the nature of the analysis, we defined two kinds of quality-flag selection:
\begin{itemize}
\item For imaging, to select events with good angular reconstruction, we used the flags {\small XMMEA\_EM} and {\small XMMEA\_EP} for the MOS and pn cameras, respectively.
\item For the spectrum analysis, we chose events with {\small FLAG=XMMEA\_SM} for MOS cameras (good energy 
reconstruction) and with {\small FLAG}=0 for the pn camera. 
\end{itemize}

The particle background was derived from filter-wheel closed (FWC) observations that were compiled until revolution about 1600. To be consistent, we applied exactly the same event selection criteria to both data and FWC files. We checked that even if the particle flux has increased significantly between 2000 and 2009, its spectrum and spatial repartition have not significantly changed during that period. We used the count rates between 10 and 12~keV for MOS and between 12 and 14~keV for pn to normalize the FWC background level to that of our observations. Regions with bright sources in the observations have been excluded to calculate the normalization factor. Finally, the {\small EVIGWEIGHT} task was used to correct vignetting effects (\cite{Pratt2007}). 

\subsection{Maps generation}

\begin{figure*}
\centering
\includegraphics[scale=0.45]{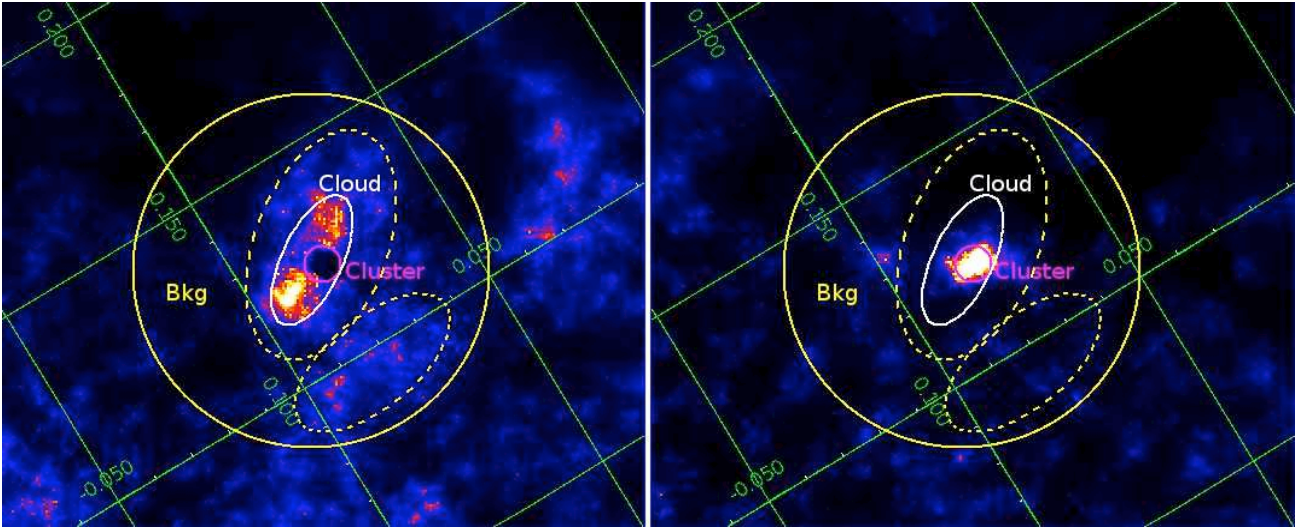}
\caption{{\it XMM-Newton}/EPIC continuum-subtracted Fe \kalpha emission line maps of the Arches cluster region at 6.4~keV ({\it left panel}) and 6.7~keV ({\it right panel}). The images have been adaptively smoothed at a signal-to-noise of 20. The magenta circle indicates the region ("Cluster") used to characterize the Arches cluster X-ray emission, which shows strong Fe \kalpha emission at 6.7 keV. The region inside the white ellipse but outside the magenta circle indicates the region ("Cloud") used for spectral extraction to characterize the bright 6.4~keV regions surrounding the Arches cluster. The region inside the yellow circle but outside the two dashed ellipses shows the local background used for the spectral analysis. The axes of the maps (in green) indicate Galactic coordinates in degrees. North is up and east to the left.}
\label{Fig15}
\end{figure*}

For each observation and instrument, we produced count images ({\small EVSELECT} task) in two energy bands (6.3--6.48~keV and 6.564--6.753~keV), which are dominated by the Fe \kalpha lines at 6.4~keV from neutral to low-ionized atoms and at 6.7~keV from a hot thermally-ionized plasma. For each energy band, observation, and instrument, the normalized particle background image derived from FWC observations was subtracted from the count image. The particle background events were rotated beforehand so as to match the orientation of each instrument for each observation. For each energy band, observation, and instrument, an exposure map was generated ({\small EEXMAP} task) taking the different efficiencies of each instrument into account. 

To produce line images of the Fe \kalpha emission at 6.4 and 6.7~keV, the continuum under the line needs to be subtracted.
For that, we produced a map in the 4.17--5.86~keV energy band, which is dominated by the continuum emission, with the same procedure as before. To determine the spectral shape of the continuum in the 4--7~keV band, the spectrum of the {\it cloud} region (representative region defined in Fig.~\ref{Fig15} and Table~\ref{Table2}) was fitted by a power law and Gaussian functions to account for the main emission lines. The continuum map was then renormalized to the power-law flux in the considered energy band and subtracted from the corresponding energy band image.

The background-subtracted and continuum-subtracted count images and the exposure maps were then merged using the {\small EMOSAIC} task to produce images of the entire observation set. The resulting count images were adaptively smoothed using the {\small ASMOOTH} task with a signal-to-noise ratio of 20. The count rate images were obtained by dividing the smoothed count images by the smoothed associated exposure map. The template of the smoothing of the count image was applied to the associated merged exposure map.

Figure~\ref{Fig15} shows the resulting Fe \kalpha line maps at 6.4 and 6.7~keV of the Arches cluster region.
The star cluster exhibits strong Fe \kalpha emission at 6.7~keV. Bright Fe \kalpha 6.4~keV structures are observed around the Arches cluster.

\begin{table}[htbp]
\begin{center}
\caption{Definition of the spectral extraction regions.}
\label{Table2} 
\begin{tabular}{|l|c|c|l|l|}
\hline
Region & RA (J2000)            & Dec (J2000)       & Shape  & Parameters  \\
\hline
\hline
Cluster & 17\hr45\mn50.3\Sec  & -28\degre49'19" & circle   & 15"  \\
\hline 
Cloud   & 17\hr45\mn51.0\Sec  & -28\degre49'16" & ellipse & 25", 59", 155\degre \\
excl.     & 17\hr45\mn50.3\Sec  & -28\degre49'19" & circle   & 15"  \\
\hline
Bkg      & 17\hr45\mn51.0\Sec  & -28\degre49'25"  & circle   & 148" \\
excl.     & 17\hr45\mn47.2\Sec  & -28\degre50'42"  & ellipse & 37", 78", 130\degre \\
excl.     & 17\hr45\mn50.4\Sec  & -28\degre49'03"  & ellipse & 55", 100", 160\degre \\
\hline
\end{tabular}
\end{center}
\tablefoot{
This Table provides the center position, circle radius and for ellipses, minor, and major axes, as well as angle (counter clockwise from straight up). ``Bkg'' means background and ``excl." indicates the zones of exclusion.
}
\end{table}

\subsection{Spectrum extraction}

To characterize the properties of the emission of the Arches cluster and its surroundings, we defined two regions, which are shown in Fig.~\ref{Fig15}. The region called ``Cluster'' corresponds to the Arches cluster and exhibits strong Fe \kalpha emission at 6.7~keV. The region called ``Cloud'' corresponds to the bright Fe \kalpha 6.4 keV emission structures surrounding the Arches cluster. Table~\ref{Table2} provides the coordinates of these regions.

In 9 of the 22 relevant observations, the Arches region lies on the out-of-service CCD6 of MOS 1, reducing the available observation time by more than a factor 2 compared to MOS 2 (see Table~\ref{Table1}). Regarding the pn camera, the cloud and cluster regions suffer from the presence of dead columns in a number of observations. We thus estimated the spatial coverage of each region for each pn observation (see Table~\ref{Table1}). After several tests, we chose for the spectral analysis to keep only observations with a pn spatial coverage greater than 85\%. For all these selected observations, the MOS spatial coverage of each region was greater than 85\% (see Table~\ref{Table1}). 

Table~\ref{Table1} summarizes the final total available exposure time by instrument for spectral analysis. With the 22 observations, we obtained 317~ks for MOS~1 and 652~ks for MOS~2. For the pn spectral analysis,  we obtained 276 and 315~ks on the cloud and cluster regions, respectively.

For each region, the particle background spectrum was estimated from the FWC observations in the same detector region.
The astrophysical background around the Arches cluster shows spatial structures and, notably, an increase towards the Galactic plane. After several tests, we concluded that the most representative local background for the Arches region
was that of the region encircling the Arches, but avoiding the zones emitting at 6.4 keV. The background region is defined in Table~\ref{Table2} and shown in Fig.~\ref{Fig15} (yellow circle with the exclusion of the two dashed-line ellipses). To subtract the particle and astrophysical background from the spectra, we used the method of double subtraction described in \cite{arnaud02}. The ancillary and redistribution matrix function response files were generated with the SAS {\small ARFGEN} and {\small RMFGEN} tasks, respectively. 

The spectra from individual observations of the same region were then merged for each instrument and rebinned to achieve a signal-to-noise per bin of 3 $\sigma$. 

\section{Variability of the 6.4 keV line}

   \begin{figure}
   \centering
   \includegraphics[width=7.5cm]{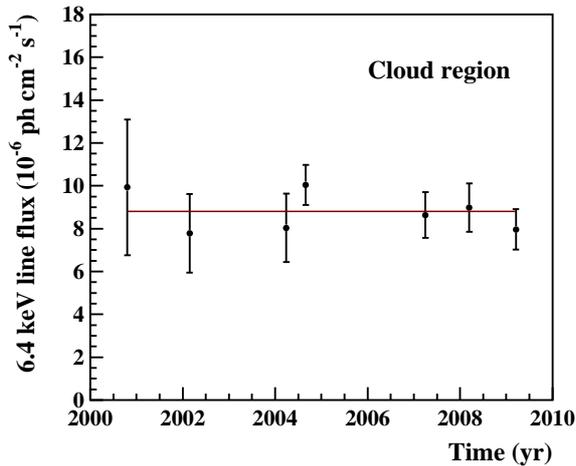}
      \caption{Lightcurve of the 6.4 keV Fe \kalpha line flux arising from a large region around the Arches 
      cluster (the region labeled ''Cloud" in Fig.~\ref{Fig15}). The red horizontal line shows the best fit
      with a constant flux. 
	}
         \label{Fig16}
   \end{figure}

The temporal variability of the 6.4~keV line flux is a key diagnostic for deciphering the origin of the line (see, for 
example, Ponti et al. \cite{pon10}). To study this aspect, we combined spectra extracted from the cloud region to 
obtain a sampling of the emission at seven epochs: September 2000 (2 observations), February 2002 (1 observation), 
March 2004 (2 observations), August/September 2004 (2 observations), March/April 2007 (3 observations), March 2008 
(2 observations), and April 2009 (3 observations). This sampling is similar to the one recently used by Capelli et 
al. (\cite{cap11b}), except that, for an unknown reason, these authors did not included the September 2000 
epoch in their analysis.  

To measure the intensity of the neutral or low-ionization Fe \kalpha line at each epoch, we modeled the X-ray 
emission from the cloud region as the sum of an optically thin, ionization equilibrium plasma (APEC, Smith et al. 
\cite{smi01}), a power-law continuum, and a Gaussian line at $\sim$6.4~keV. These three components were subject to 
a line-of-sight photoelectric absorption so as to account for the high column density of the foreground material. 
We used the X-ray spectral-fitting program XSPEC\footnote{\url{http://heasarc.nasa.gov/xanadu/xspec/}} to fit this 
model simultaneously to EPIC MOS and pn spectra between 1.5 and 10~keV. More details on the fitting procedure will 
be given in the next section. All the fits were satisfactory and gave reduced $\chi^2 \sim 1$. 

The photon fluxes in the 6.4~keV line thus determined are shown in Fig.~\ref{Fig16}. The best fit of a constant 
flux to these data is satisfactory, giving a $\chi^2$ of 3.3 for six degrees of freedom (dof). The significance of 
a variation in the line flux is then only of 0.3$\sigma$\footnote{Noteworthy is that these results were obtained without 
taking the systematic error in the effective area of the EPIC camera into account. This error is estimated to be 
7\% for on-axis sources and to increase with off-axis angle (see the {\it XMM-Newton} Calibration 
Technical Note CAL-TN-0018.pdf at \url{http://xmm.vilspa.esa.es/external/xmm_sw_cal/calib/documentation.shtml}).}. 
The best-fit mean flux is $F_{\rm 6.4~keV}=(8.8\pm0.5)\times 10^{-6}$~ph~cm$^{-2}$~s$^{-1}$. 

The fact that the intensity of the 6.4~keV line emitted from the vicinity of the Arches cluster is consistent with 
being constant is in good agreement with the previous work of Capelli et al. (\cite{cap11b}). We note, however,
that the line fluxes obtained in the present work are systematically lower by $\sim$20\%. This is 
attributable to a difference in the background modeling: whereas we used a broad region as close as possible to 
the Arches cluster to subtract the astrophysical background prior to the spectral fitting (Fig.~\ref{Fig15}), 
Capelli et al. included the background as a component within the fitting. 

\section{X-ray spectral analysis of time-averaged spectra}

\begin{table*}
 \centering
  \caption{Spectral analysis of the X-ray emission from the Arches star cluster 
  and associated cloud region with standard XSPEC models.}
  \label{Table3}
  \begin{tabular}{|c|c|ccc|c|}
   \hline
 & (Unit) & \multicolumn{3}{c|}{star cluster} & cloud region \\
 & & model 1 & model 2 & model 3 & model 1 \\
   \hline
   \hline
 \NH(2) & ($10^{22}${ \NHUNIT}) & -- & = \NH(1) & 8.3$^{+0.8}_{-1.0}$ & -- \\
 $kT$(2)  & (keV) & -- & 0.27$\pm$0.04 & 0.92$^{+0.14}_{-0.15}$ & -- \\
 $Z/Z_\odot$  &  & -- & 1.7 (fixed) & 1.7 (fixed) & -- \\
 $I_{kT}{\rm (2)}$ & (see notes below) & -- & 1100$^{+1500}_{-600}$ & 13$^{+7}_{-5}$ & -- \\
   \hline
 \NH(1) & ($10^{22}${ \NHUNIT}) & 9.5$\pm$0.3 & 12.0$^{+0.6}_{-0.7}$ & 12.8$^{+1.3}_{-1.0}$ & 11.3$^{+1.9}_{-1.3}$ \\
   \hline
 $kT$(1)  & (keV) & 1.79$^{+0.06}_{-0.05}$ & 1.61$^{+0.08}_{-0.05}$ & 1.78$^{+0.15}_{-0.10}$ & 2.2$^{+1.0}_{-0.5}$ \\
 $Z/Z_\odot$  &  & 1.7 (fixed) & 1.7 (fixed) & 1.7 (fixed) & 1.7 (fixed) \\
 $I_{kT}{\rm (1)}$ & (see notes below) & 20.4$\pm$1.8 & 30$\pm$4 & 23$^{+4}_{-5}$ & 5.2$^{+4.9}_{-2.4}$ \\
   \hline
 $E_{\rm 6.4~keV}$ & (keV) & 6.41$\pm$0.02 & 6.40$\pm$0.02 & 6.40$\pm$0.02 & 6.409$\pm$0.005 \\
 $F_{\rm 6.4~keV}$ & (10$^{-6}$~ph~cm$^{-2}$~s$^{-1}$) & 1.2$\pm$0.3 & 1.3$\pm$0.3 & 1.3$\pm$0.3 & 8.7$^{+0.5}_{-0.6}$ \\
   \hline
 $\Gamma$ &  & 0.7$\pm$0.4 & 0.4$^{+0.5}_{-0.6}$ & 0.8$^{+0.6}_{-0.7}$ & 1.6$^{+0.3}_{-0.2}$ \\
 $I_{\rm p.l.}$ & (10$^{-5}$~cm$^{-2}$~s$^{-1}$~keV$^{-1}$) & 1.3$^{+1.4}_{-0.7}$ & 0.62$^{+1.15}_{-0.35}$ & 1.3$^{+2.8}_{-0.9}$ & 16$^{+9}_{-6}$ \\
   \hline
 EW$_{\rm 6.4~keV}$ & (keV) & 0.4$\pm$0.1 & 0.4$\pm$0.1 & 0.4$\pm$0.1 & 1.2$\pm$0.2 \\
   \hline
 $\chi^2$/dof &  & 1222/978 & 1152/976 & 1129/975 & 560/491 \\
   \hline
  \end{tabular}
   \tablefoot{Model 1: WABS$\times$(APEC + Gaussian + powerlaw); model 2: WABS$\times$(APEC + APEC + Gaussian + powerlaw);
  model 3: WABS$\times$APEC + WABS$\times$(APEC + Gaussian + powerlaw).
  \NH: absorption column density. $kT$, $Z/Z_\odot$, and $I_{kT}$: temperature, metallicity 
   relative to solar, and normalization of the APEC thermal plasma ($I_{kT}$ is in unit of $10^{-18} 
   \int n_{\rm e} n_{\rm H} dV / (4\pi D^2) $, where $n_{\rm e}$ and $n_{\rm H}$ are the electron 
   and proton number densities (cm$^{-3}$) and $D$ the distance to the source in cm). 
   $E_{\rm 6.4~keV}$ and $F_{\rm 6.4~keV}$: centroid energy and flux of the neutral or low-ionization Fe \kalpha line. 
   $\Gamma$ and $I_{\rm p.l.}$: index and normalization at 1~keV of the power-law component. 
   EW$_{\rm 6.4~keV}$: EW of the 6.4 keV line with respect to the power-law continuum. $\chi^2$/dof: 
   $\chi^2$ per degree of freedom.}
\end{table*}

\begin{figure*}
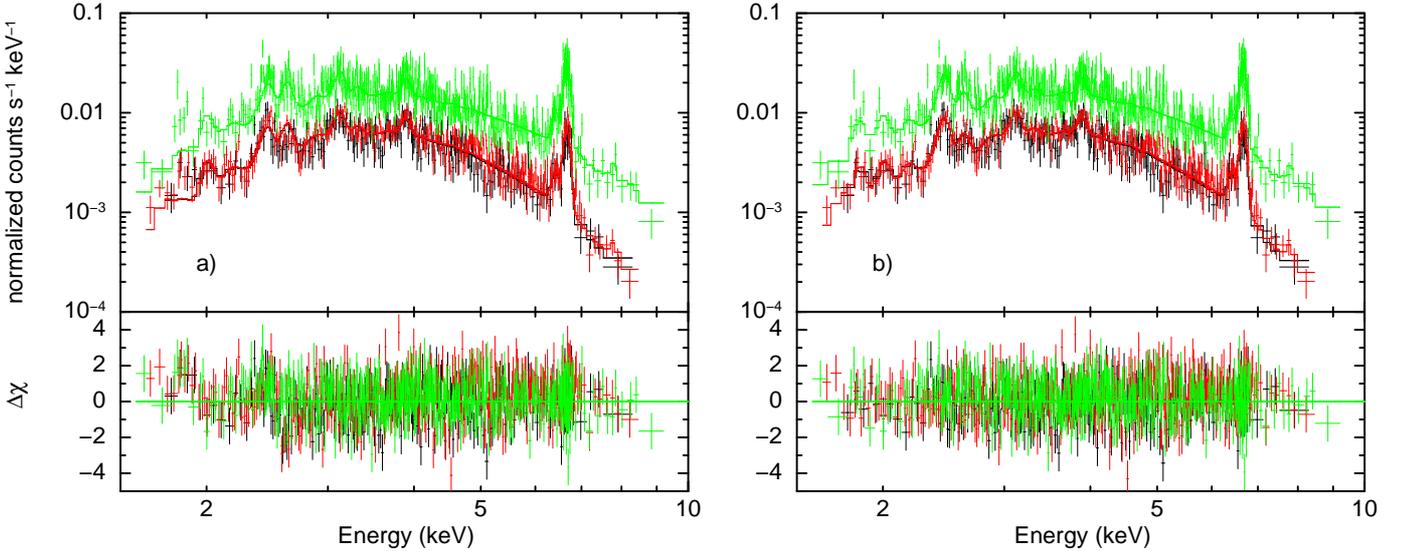

\hspace{0cm}
\begin{minipage}{0.46\linewidth}
\includegraphics[scale=0.4,angle=-90]{Cluster_OneAPEC.eps}
\end{minipage}
\hfill
\begin{minipage}{0.46\linewidth}
\includegraphics[scale=0.4,angle=-90]{Cluster_TwoAPEC_2.eps}
\end{minipage}
\caption{X-ray spectra of the Arches cluster as measured in the MOS1 (black), MOS2 (red), and pn (green) cameras aboard 
{\it XMM-Newton}, compared to {\bf a)} a model with only one thermal plasma component (model 1 in Table~\ref{Table3}) and 
{\bf b)} a model with two thermal plasma components (model 3 in Table~\ref{Table3}). The lower panels show the associated 
residuals in terms of standard deviations. The second plasma of temperature $kT = 0.9$~keV accounts for a significant 
emission in the He-like Si and S \kalpha lines at 1.86 and 2.46~keV, respectively.}
\label{Fig17}
\end{figure*}

We again used XSPEC to fit various models to time-averaged spectra extracted from the two 
source regions shown in Fig.~\ref{Fig15}. The fits were performed simultaneously 
on the stacked MOS1, MOS2, and pn spectra, but we allowed for a variable cross-normalization factor between 
the MOS and pn data. Independent of the fitting model, we found very good agreement between the MOS and pn 
cameras, to better than 1\% for the data extracted from the cluster region and to 4--5\% for the data extracted 
from the cloud region. These factors are consistent with the residual uncertainty in the flux cross-calibration of the 
EPIC cameras (Mateos et al. \cite{mat09}).

\subsection{X-ray emission from the star cluster}

We first modeled the emission of the cluster region as the sum of an APEC plasma component and a nonthermal 
component represented by a power-law continuum and a Gaussian line at $\sim$6.4~keV. The centroid energy of 
the Gaussian line was allowed to vary, but the line width was fixed at 10~eV. All the emission components were 
subject to a line-of-sight photoelectric absorption (WABS model in XSPEC). The best-fit results obtained with this 
model (called model 1 in the following) are reported in Table~\ref{Table3} and the corresponding spectra shown 
in Fig.~\ref{Fig17}a. In this table and in the following discussion, all the quoted errors are at the 90\% 
confidence level.

This fitting procedure did not allow us to reliably constrain the 
metallicity of the X-ray emitting plasma. Indeed, the best fit was obtained for a super-solar metallicity 
$Z > 5\,Z_\odot$, which is not supported by other observations. Such an issue has 
already been faced in previous analyses of the X-ray emission from the Arches cluster region. Thus, 
Tsujimoto et al. (\cite{tsu07}) fixed the plasma metallicity to be solar in their analysis of {\it Suzaku} 
data, whereas Capelli et al. (\cite{cap11a,cap11b}) adopted $Z = 2\,Z_\odot$ in their analysis of 
{\it XMM-Newton} data. With the {\it Chandra X-ray Observatory}, Wang et al. (\cite{wan06}) were able to 
resolve three bright point-like X-ray sources in the core of the Arches cluster -- most likely colliding 
stellar wind binaries -- and study them individually. These sources were all modeled by an 
optically thin thermal plasma with a temperature of $\sim$1.8--2.5~keV and a metallicity  
$Z/Z_\odot = 1.8^{+0.8}_{-0.2}$. In our analysis, we fixed the metallicity of the thermal plasma in the 
cluster region to be $1.7\,Z_\odot$, which is the best-fit value that we were able to obtain for the cloud 
region using the LECR ion model developed in this paper to account for the nonthermal emission (see 
Sect.~5.2 below). The adopted metallicity is also consistent with the results of Wang et al. (\cite{wan06}).

\begin{figure}
   \centering
   \includegraphics[width=7.5cm]{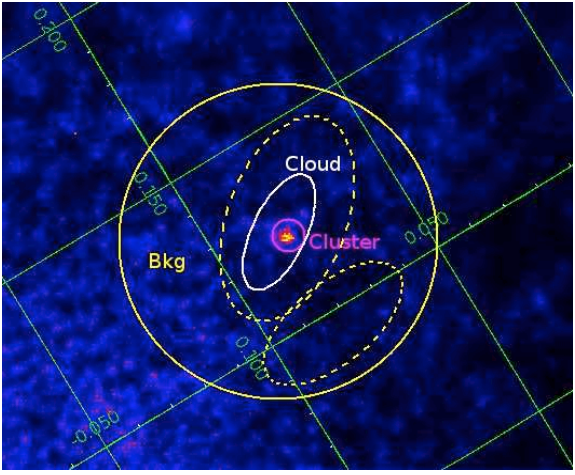}
      \caption{Same as Fig.~\ref{Fig15} but for the He-like Si \kalpha line at 1.86~keV. 
              }
         \label{Fig17prim}
   \end{figure}

Model 1 gives a good fit to the data from the cluster region above $\sim$3~keV. In particular, the detection
of the neutral or low-ionization Fe K$\alpha$ line is significant (see Table~\ref{Table3}). But the fit is 
poorer below 3~keV, because the data shows clear excesses of counts above the model at $\sim$1.85 and $\sim$2.45~keV 
(see Fig.~\ref{Fig17}a). These features most likely correspond to the K$\alpha$ lines from He-like Si and S, 
respectively. We checked that this excess emission is not due to an incomplete background subtraction 
by producing a Si K$\alpha$ line image in the energy band 1.76--1.94~keV. To estimate the contribution 
of the continuum under the Si line, we first produced a count map in the adjacent energy band 2.05--2.15~keV 
and then normalized it to the expected number of continuum photons in the former energy range. The 
normalization factor was obtained from a fit to the EPIC spectra of the cluster region by model 1 plus two 
Gaussian functions to account for the Si and S \kalpha lines. The resulting map in the Si line shows 
significant excess emission at the position of the Arches cluster (Fig~\ref{Fig17prim}).

To account for the presence of the He-like Si and S \kalpha lines in the X-ray spectrum of the cluster, we 
included in the fitting model a second APEC component subject to the same photoelectric absorption as the other 
components (model 2). The quality of the fit significantly improves with this additional thermal plasma component 
($\chi^2$=1152 for 976 degrees of freedom), whose best-fit temperature is $kT=0.27\pm0.04$~keV 
(Table~\ref{Table3}). However, the absorption-corrected intrinsic luminosity of this plasma is found to be quite 
high in the soft X-ray range: $L_{\rm int}(0.4-1~{\rm keV})\approx 2.3\times 10^{36}$~erg~s$^{-1}$, assuming the 
distance to the GC to be $D=8$~kpc (Ghez et al. \cite{ghe08}). 

In a third model, we let the X-ray emission from the plasma of lower temperature be absorbed by a 
different column density than the one absorbing the X-rays emitted from the other components. It allows us to 
further improve the fit to the data ($\chi^2$=1129 for 975 degrees of freedom, Table~\ref{Table3}; see also 
Fig.~\ref{Fig17}b for a comparison of this model to the data). We then found $kT=0.92^{+0.14}_{-0.15}$~keV and 
$L_{\rm int}(0.4-1~{\rm keV})\approx 2.0\times 10^{34}$~erg~s$^{-1}$ for the lower temperature plasma. The 
origin of this thermal component, which was not detected in previous X-ray observations of the Arches 
cluster, is discussed in Sect.~6.1 below.  

As can be seen from Table~\ref{Table3}, the addition of a second APEC component in the fitting model of the 
star cluster emission increases the absorbing column density \NH(1) significantly. It also has some impact 
on the temperature of the hotter plasma and on the index of the power-law component, but not on the properties 
of the 6.4~keV line. 

\subsection{X-ray emission from the cloud region}

We used model 1 to characterize the X-ray emission from the cloud region, except that we also included a 
Gaussian line at 7.05~keV (fixed centroid energy) to account for the neutral or low-ionization Fe K$\beta$ line. 
The Fe K$\beta$/\kalpha flux ratio was imposed to be equal to 0.13 (Kaastra \& Mewe \cite{kaa93}). We checked 
that including a second thermal plasma component (model 2 or 3) is not required for this region, as it does 
not improve the quality of the fit. As before, we fixed the metallicity of the emitting plasma to be $1.7\,Z_\odot$. 
The best-fit temperature, $kT=2.2^{+1.0}_{-0.5}$~keV, is marginally higher than the one of the high-temperature 
plasma emanating from the cluster region. 

\begin{table}
 \centering
  \caption{Spectral analysis of the X-ray emission from the cloud 
  region with LECR electron and ion models.}
  \label{Table4}
  \begin{tabular}{|c|c|c|c|}
   \hline
 & (Unit) & LECR electrons & LECR ions \\
   \hline
   \hline
 \NH & ($10^{22}${ \NHUNIT}) & 11.9$^{+1.3}_{-1.4}$ & 12.2$^{+1.4}_{-1.6}$ \\
   \hline
 $kT$  & (keV) & 1.9$^{+0.6}_{-0.3}$ & 2.0$^{+0.7}_{-0.3}$ \\
 $Z/Z_\odot$  &  &  $>3.1$ & 1.7$\pm 0.2$ \\
 $I_{kT}$ & (see notes below) & 3.5$^{+2.4}_{-1.6}$ & 7.0$^{+4.0}_{-3.1}$ \\
   \hline
 $\Lambda$ & (H-atoms cm$^{-2})$ & $5 \times 10^{24}$ (fixed) & $5 \times 10^{24}$ (fixed) \\
 $s$ &  & $>2.5$ & 1.9$^{+0.5}_{-0.6}$ \\
 $E_{\rm min}$ & (keV) or (keV/n) & $< 41$ & 10$^4$ (fixed) \\
 $N_{\rm LECR}$ & (10$^{-8}$~erg~cm$^{-2}$~s$^{-1}$) & 5.0$^{+7.4}_{-1.5}$ & 5.6$^{+0.7}_{-0.3}$ \\
   \hline
 $\chi^2$/dof &  & 558/492 & 558/493 \\
   \hline
  \end{tabular}
   \tablefoot{XSPEC model: WABS$\times$(APEC + LECR$p$), where $p$ stands for electrons or ions.
   \NH, $kT$, $Z/Z_\odot$, and $I_{kT}$: as in Table~\ref{Table3}. $\Lambda$,
   $s$, $E_{\rm min}$, and $N_{\rm LECR}$: LECR path length, source spectrum index, minimum energy, and 
   model normalization. By definition $dW/dt=4\pi D^2 N_{\rm LECR}$ is the power injected in the interaction region 
   by primary CR electrons or protons of energies between $E_{\rm min}$ and $E_{\rm max}=1$~GeV ($D$ is the distance 
   to the source).}
\end{table}

\begin{figure*}
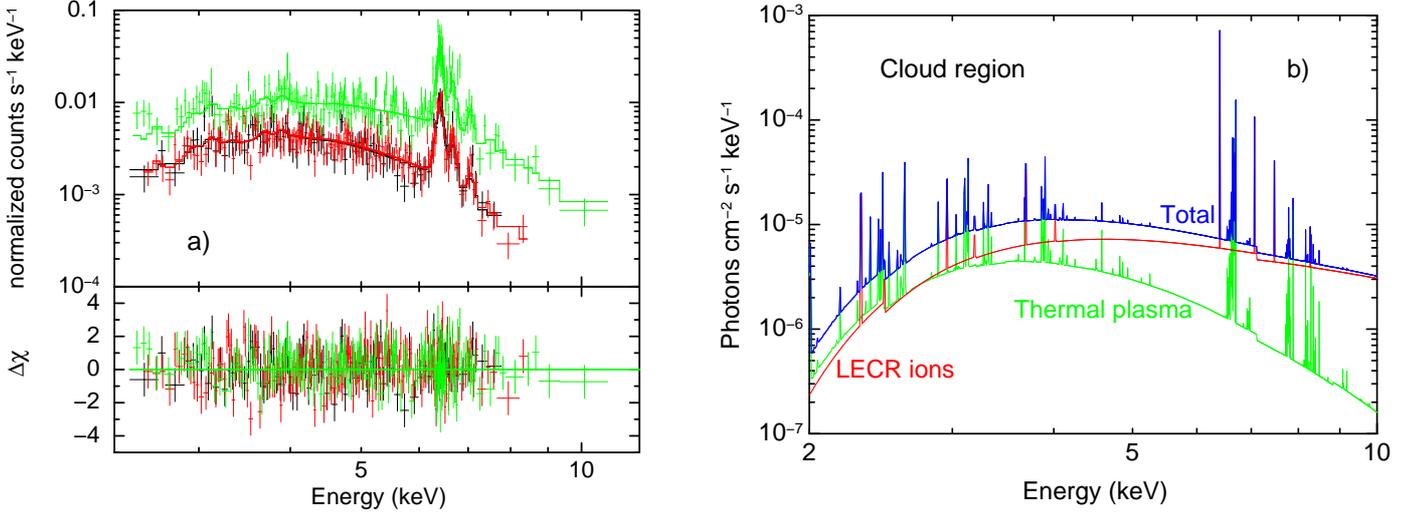

\hspace{0cm}
\begin{minipage}{0.48\linewidth}
\includegraphics[scale=0.37,angle=-90]{Cloud_LECRp.eps}
\end{minipage}
\hfill
\begin{minipage}{0.48\linewidth}
\includegraphics[scale=0.35,angle=-90]{Model_Cloud_LECRp.eps}
\end{minipage}
\caption{{\bf a)} X-ray spectra of the cloud region as measured in the {\it XMM-Newton} cameras and the best-fit
spectral model assuming that the emission comes from a combination of a collisionally ionization equilibrium plasma 
(APEC model) and a nonthermal component produced by interactions of LECR ions with the cloud constituents (see 
Table~\ref{Table4}); {\bf b)} model components.}
\label{Fig18}
\end{figure*}

We now compare the characteristics of the prominent nonthermal emission of the cloud region with the model 
predictions discussed in Sect.~2. In the LECR electron model, the measured value 
$\Gamma = 1.6^{+0.3}_{-0.2}$ would only be expected for low values of the CR minimum energy $E_{\rm min} \lsim 100$~keV 
and for relatively soft source spectra with $s \gsim 2.5$ (see Fig.~\ref{Fig7}). But for these CR spectrum parameters 
the neutral Fe K$\alpha$ line is predicted to be relatively weak, EW$_{\rm 6.4~keV} < 0.4\times (Z/Z_\odot)$~keV 
(see Fig.~\ref{Fig6}). Thus, it would require an ambient Fe abundance $\gsim 3$ times the solar value to account for 
the measured EW of $1.2 \pm 0.2$~keV. The measured properties of the nonthermal component emitted from the cloud 
region thus appear to be hardly compatible with the predictions of the LECR electron model. 

On the other hand, the measured values of $\Gamma$ and EW$_{\rm 6.4~keV}$ for the cloud region seem to be compatible 
with the LECR ion model. The measured power-law slope can be produced in this model with any spectral index $s \sim 1.5$--2, 
provided that the CR path length $\Lambda > 10^{24}$~cm$^{-2}$ (Figs.~\ref{Fig12} and \ref{Fig14}). We then expect 
EW$_{\rm 6.4~keV} \sim (0.6$--$1)\times (Z/Z_\odot)$~keV (Figs.~\ref{Fig11} and \ref{Fig13}), which would be in 
good agreement with the measured EW for an ambient metallicity $Z \lsim 2 \,Z_\odot$.  

To further study the origin of the prominent nonthermal emission of the cloud region, we created LECR electron and 
ion models that can be used in the XSPEC software. For this purpose, a total of 70875 spectra were calculated for each 
model by varying the four free parameters of the models in reasonable ranges. The calculated spectra were then gathered 
in two FITS files that can be included as external models in XSPEC\footnote{These models are available upon 
request to the authors.}. We then fitted the stacked spectra of the cloud region by the XSPEC model 
WABS$\times$(APEC + LECR$p$), where $p$ stands for electrons or ions. The best-fit results obtained with both models 
are given in Table~\ref{Table4}.

In this spectral fitting, we allowed for a variable metallicity of the nonthermal X-ray production region (i.e. the 
parameter $Z$ of the LECR$p$ models), but we imposed this parameter to be equal to the metallicity of the thermal 
plasma. Since both fits did not usefully constrain the path length of the LECRs in the interaction region, we fixed 
$\Lambda=5\times 10^{24}$~cm$^{-2}$ for both models, which, as discussed in Appendix~A, is a typical value for 
nonrelativistic protons propagating in massive molecular clouds of the GC environment (see Eq.~(\ref{eq6})). 
As anticipated, the LECR electron model cannot satisfactorily account for the data, because the best fit is 
obtained for too high a metallicity ($Z > 3.1 \,Z_\odot$; limit at the 90\% confidence level) and a low CR minimum 
energy ($E_{\rm min} < 41$~keV). This conclusion is independent of the adopted value of $\Lambda$. 

On the other hand, the data can be characterized well by a thin plasma component plus an LECR ion model. In particular, 
the best-fit metal abundance $Z/Z_\odot = 1.7\pm 0.2$ is in good agreement with previous works (Wang et al. \cite{wan06}). 
The best-fit CR spectral index is $s=1.9^{+0.5}_{-0.6}$. For such a relatively hard CR source spectrum, one can see from  
Figs.~\ref{Fig13}a and \ref{Fig14} that the nonthermal X-ray emission produced by LECR ions only weakly depends on the 
CR minimum energy $E_{\rm min}$. Accordingly, the fit did not constrain this parameter, which was finally fixed at 
$E_{\rm min}=10$~MeV~nucleon$^{-1}$. As discussed in Appendix~A, the process of CR penetration into molecular clouds is not 
understood well, such that $E_{\rm min}$ is loosely constrained from theory. This parameter has an effect, however, on 
the power injected by the primary CRs in the X-ray production region (see Fig.~\ref{Fig13}b). Thus, for 
$E_{\rm min}=1$~MeV~nucleon$^{-1}$ (resp. $E_{\rm min}=100$~MeV~nucleon$^{-1}$), the best-fit normalization of the LECR 
ion model is $N_{\rm LECR}=(7.4^{+8.3}_{-1.5})\times 10^{-8}$~erg~cm$^{-2}$~s$^{-1}$ 
(resp. $(3.1^{+2.8}_{-0.2})\times 10^{-8}$~erg~cm$^{-2}$~s$^{-1}$). The corresponding power injected by LECR protons in 
the cloud region ($dW/dt=4\pi D^2 N_{\rm LECR}$) lies in the range (0.2--1)$\times 10^{39}$~erg~s$^{-1}$ (with $D=8$~kpc).  

The best-fit model obtained with the LECR ion component is compared to the data of the cloud region in Fig.~\ref{Fig18}a, 
and the corresponding theroretical spectrum is shown in Fig.~\ref{Fig18}b. The latter figure exhibits numerous 
lines arising from both neutral and highly-ionized species, which could be revealed by a future instrument 
having an excellent sensitivity and energy resolution. 

\section{Origin of the detected radiations}

We have identified three distinct components in the X-ray spectra extracted from the cluster region: an optically thin 
thermal plasma with a temperature $kT$$\sim$1.6--1.8~keV, another plasma of lower temperature ($kT$$\sim$0.3~keV in 
model 2 or $\sim$0.9~keV in model 3), and a relatively weak nonthermal component characterized by a hard 
continuum emission and a line at 6.4~keV from neutral to low-ionized Fe atoms (EW$_{\rm 6.4~keV}=0.4\pm 0.1$~keV). The 
X-ray radiation arising from the cloud region is also composed of a mix of a thermal and a nonthermal component, but 
the 6.4 keV Fe K$\alpha$ line is much more intense from there, with a measured EW of $1.2\pm 0.2$~keV. 

\subsection{Origin of the thermal X-ray emissions}

The thermal component of temperature $kT$$\sim$1.6--1.8~keV detected from the star cluster most likely arises from several 
colliding stellar wind binaries plus the diffuse hot plasma of the so-called cluster wind. Wang et al. (\cite{wan06}) 
find with the {\it Chandra} telescope three point-like sources of thermal emission with $ kT$$\sim$1.8--2.5~keV 
embedded in a spatially extended emission of similar temperature. Capelli et al. (\cite{cap11a}) have recently found with 
{\it XMM-Newton} that the bulk of the X-ray emission from the Arches cluster can be attributed to an optically thin 
thermal plasma with a temperature $kT$$\sim$1.7~keV. The diffuse thermal emission from the cluster is thought to be 
produced by the thermalization of massive star winds that merge and expand together. The expected temperature of such 
a cluster wind is consistent with the temperature of the hot thermal component identified in this and previous works 
(see Capelli et al. \cite{cap11a} and references therein). 

The plasma with $kT$$\sim$1.6--1.8~keV is at the origin of the He-like Fe K$\alpha$ line at 6.7~keV. The 
corresponding map generated in the present work (Fig.~\ref{Fig15}, {\it right panel}) is in good agreement with the 
{\it Chandra} observations. Wang et al. (\cite{wan06}) suggest that the observed elongation of this 
emission in the east-west direction reflects an ongoing collision of the Arches cluster with a local molecular 
cloud traced by the CS emission. As discussed by Wang et al. (\cite{wan06}), this collision may help in 
explaining the spatial confinement of this hot plasma.

The second thermal component of temperature $kT$$\sim$0.3~keV (model 2) or $kT$$\sim$0.9~keV (model 3) was not 
detected in previous X-ray observations of the Arches cluster. An optically thin thermal plasma of temperature 
$kT$$\sim$0.8~keV was reported by Yusef-Zadeh et al. (\cite{yus02b}) from {\it Chandra} observations, but not by 
subsequent X-ray observers (Wang et al. \cite{wan06}, Tsujimoto et al. \cite{tsu07}, Capelli et al. \cite{cap11a}). 
However, the high-quality spectral data obtained in the present work reveal that a single APEC thermal plasma 
model cannot account simultaneously for the observed lines at $\sim$1.85, $\sim$2.45, and 6.7~keV, which arise from 
He-like Si, S, and Fe atoms, respectively. The map at $\sim$1.85~keV clearly shows that the star cluster 
significantly emits at this energy (Fig.~\ref{Fig17prim}). The required additional plasma component is subject to a 
high interstellar absorption: $N_{\rm H} \approx 1.2\times 10^{23}$ in model 2 and $8.3\times 10^{22}$~H~cm$^{-2}$
in model 3 (Table~\ref{Table3}). It shows that the emitting plasma is located in the Galactic 
center region and not in the foreground. 

The temperature of the second thermal component suggests that this emission could be due to a collection of 
individual massive stars in the cluster. Single hot stars with spectral types O and early B are known to emit 
significant amounts of thermal X-rays with a temperature $kT$ in the range 0.1--1~keV and a typical luminosity in soft 
X-rays $L_X(0.4-1~{\rm keV})  \sim 1.5 \times 10^{-7} L_{\rm bol}$ (Antokhin et al. \cite{ant08}; G\"udel \& Naz\'e 
\cite{gud09}). Here, $L_{\rm bol}$ is the bolometric luminosity of the star. The total bolometric luminosity 
of the Arches cluster is $\sim 10^{7.8}$~$L_\odot$ and most of it is contributed by early B- and O-type stars, some of 
which have already evolved to the earliest Wolf-Rayet phases (Figer et al. \cite{fig02}). Then, a total soft X-ray 
luminosity $L_X(0.4-1~{\rm keV}) \sim 3.6 \times 10^{34}$~erg~s$^{-1}$ can be expected from the ensemble of hot 
massive stars of the cluster. This estimate is much lower than the unabsorbed intrinsic luminosity of the 
$\sim$0.3~keV plasma found in model 2, $L_{\rm int}(0.4-1~{\rm keV})\approx 2.3\times 10^{36}$~erg~s$^{-1}$. But 
it is roughly consistent with the absorption-corrected luminosity of the $\sim$0.9~keV plasma found in model 3: 
$L_{\rm int}(0.4-1~{\rm keV})\approx 2.0\times 10^{34}$~erg~s$^{-1}$. It is not clear, however, why the latter 
component is less absorbed than the high-temperature plasma emitted from the Arches cluster 
(Table~\ref{Table3}). 

\subsection{Origin of the 6.4~keV line emission}

\subsubsection{The cloud region}

Several molecular clouds of the GC region emit the 6.4 keV line, most notably Sgr B1, Sgr B2, Sgr C, and clouds 
located between Sgr~A$^\ast$ and the Radio Arc (see Yusef-Zadeh et al. \cite{yus07}). Detections of time variability 
of the 6.4~keV line from Sgr B2 (Inui et al. \cite{inu09}), as well as from molecular clouds within $15'$ to the east 
of Sgr~A$^\ast$ (Muno et al. \cite{mun07}; Ponti et al. \cite{pon10}), are best explained 
by the assumption that the Fe K$\alpha$ line emission from these regions is a fluorescence radiation produced by the 
reprocessing of a past X-ray flare from the supermassive black hole Sgr~A$^\ast$. In this model, the variability of the 
line flux results from the propagation of an X-ray light front emitted by Sgr~A$^\ast$ more than $\sim 100$~years ago. 
The discovery of an apparent superluminal motion of the 6.4~keV line emission from the so-called ``bridge'' region provides
strong support for this model (Ponti et al. \cite{pon10}). The observed line flux variability with a timescale of a few 
years is hard to explain by a model of CR irradiation. 

In contrast to these results, the flux of the neutral or low-ionization Fe K$\alpha$ line emitted from the Arches cluster 
vicinity does not show any significant variation over more than eight years of {\it XMM-Newton} repeated observations 
performed between 2000 and 2009 (Sect.~4). Capelli et al. (\cite{cap11b}) divided the zone of 6.4~keV line emission 
around the cluster into two subregions of about one parsec scale (labeled ``N'' and ``S'' by these authors) and found 
that both subregions emit the line at a constant flux. Other regions in the central molecular zone have been observed to 
emit a steady 6.4~keV line emission during about the same period, but they generally have larger spatial extents (see, e.g., 
Ponti et al. \cite{pon10}). Thus, the spatially averaged Fe K$\alpha$ emission from Sgr B2 appears to have almost been 
constant for more than about seven years before fading away (Inui et al. \cite{inu09}; Terrier et al. \cite{ter10}), which 
is compatible with the light crossing time of the molecular cloud complex (see Odaka et al. \cite{oda11}). But recent 
observations of Sgr B2 with {\it Chandra} suggest that the overall emission of the complex at 6.4 keV is in fact composed 
of small structures that have constantly changed shape over time (Terrier et al., in preparation). 

Together with the nondetection of time variability, the poor correlation of the spatial distribution of the 6.4~keV line 
emission with that of the molecular gas also argues against any origin of the Fe line in the Arches cluster region related 
to Sgr~A$^\ast$ (Wang et al. \cite{wan06}). Lang et al. (\cite{lan01,lan02}) studied the position of the molecular clouds 
in the vicinity of the cluster by combining CS(2--1) observations with H92$\alpha$ recombination line data. The latter 
were used to trace the Arched filaments \HII\ regions, which are thought to be located at edges of molecular clouds 
photoionized by the adjacent star cluster. Lang et al. (\cite{lan01,lan02}) show that the molecular material in 
this region has a finger-like distribution and that the cluster is located in the midst of the so-called 
``$-30$~km~s$^{-1}$ cloud" complex, which extends over a region of $\sim$20~pc diameter (see also Serabyn \& G\"uesten 
\cite{ser87}). 

\begin{figure}
   \centering
   \includegraphics[width=7.8cm]{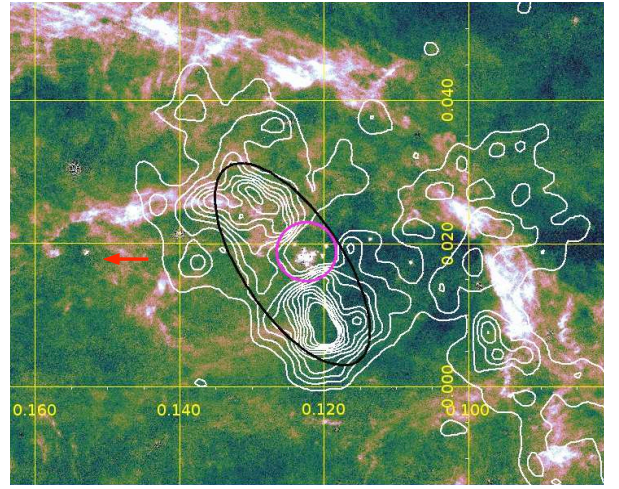}
      \caption{\textit{XMM-Newton}/EPIC continuum-subtracted 6.4-keV line intensity contours (linearly spaced between $3\times 10^{-8}$ and $1.8\times 10^{-7}$ photons~cm$^{-2}$~s$^{-1}$~arcmin$^{-2}$) overlaid with an \textit{HST}/NICMOS map in the H Paschen-$\alpha$ line (Wang et al. \cite{wan10}; Dong et al. \cite{don11}). The axes of the map indicate Galactic coordinates in degrees. The black ellipse and the magenta circle show the two regions used for spectral extraction (see Fig.~\ref{Fig15}). The red arrow illustrates the observed proper motion of the Arches cluster, which is almost parallel to the Galactic plane (Stolte et al. \cite{sto08}; Clarkson et al. \cite{cla12}). North is up and east to the left.}
         \label{Fig19}
   \end{figure}

Figure~\ref{Fig19} compares the distribution of the 6.4 keV line emission around the star cluster with a high-resolution 
image in the hydrogen Paschen-$\alpha$ (P$\alpha$) line recently obtained with the {\it Hubble Space 
Telescope}/NICMOS instrument (Wang et al. \cite{wan10}; Dong et al. \cite{don11}). The P$\alpha$ line emission is a sensitive 
tracer of massive stars --~the Arches cluster is clearly visible in this figure at Galactic coordinates $(\ell,b) \approx 
(0.122^\circ,0.018^\circ$)~-- and of warm interstellar gas photoionized by radiation from these stars. The main diffuse 
P$\alpha$-emitting features in Fig.~\ref{Fig19} are the three easternmost Arched filaments: E1 (at $0.13^\circ \lsim \ell 
\lsim 0.15^\circ$ and $b\sim0.025^\circ$), E2 (at $b\gsim0.04^\circ$), and G0.10+0.02 (running from $(\ell,b) \sim 
(0.09^\circ,0.006^\circ)$ to $(0.1^\circ,0.025^\circ)$; see also Lang et al. \cite{lan02}). The 6.4~keV line emission is 
not well correlated with the Arched filaments. The most prominent structure at 6.4 keV is concentrated in a region 
of only a few pc$^2$ surrounding the star cluster, much smaller than the spatial extent of the $-30$~km~s$^{-1}$ cloud complex. 
The origin of the faint Fe K emission at $(\ell,b) \sim (0.10^\circ,0.02^\circ$) is discusssed in the next section. This 
strongly suggests that the origin of the bright nonthermal X-ray radiation is related to the cluster itself and not to a 
distant source such as Sgr~A$^\ast$.

Assuming that the nonthermal emission from the cloud region is produced by a hard X-ray photoionization source located in 
the Arches cluster, the 4 -- 12~keV source luminosity required to produce the observed 6.4~keV line flux can be estimated 
from Sunyaev \& Churazov (\cite{sun98}):
\begin{eqnarray} 
L_X & \sim & 10^{36}~{\rm erg~s}^{-1} \times \bigg({F_{\rm 6.4~keV} \over 8.7 \times 10^{-6}{\rm~ph~cm^{-2}~s^{-1}}}\bigg) 
\nonumber \\ & & \times~ \bigg({Z \over Z_\odot}\bigg)^{-1} \bigg({N_{\rm H}^C \over 10^{23}{\rm~cm^{-2}}}\bigg)^{-1} 
\bigg({\Omega \over 0.1}\bigg)^{-1}~, 
\label{eq13p}
\end{eqnarray} 
where the distance to the GC is again assumed to be 8~kpc. Here, $F_{\rm 6.4~keV}$ is the measured 6.4~keV line flux 
(Table~\ref{Table3}), $N_{\rm H}^C$ is the column density of the line-emitting cloud, and $\Omega$ the fractional 
solid angle that the cloud subtends at the X-ray source. This quantity is called the covering factor in, e.g., Yaqoob et 
al. (\cite{yaq10}). In comparison, the unabsorbed luminosity of the cluster that we measured from the time-averaged 
{\it XMM-Newton} spectra is $L_X(4-12~{\rm keV})\approx 5\times 10^{33}$~erg~s$^{-1}$. Capelli et al. (\cite{cap11a}) 
recently detected a 70\% increase in the X-ray emission of the Arches cluster in March/April 2007. However, 
the observed X-ray luminosity of the cluster is about two orders of magnitude short of what is required for the 
fluorescence interpretation.

An alternative hypothesis is that the 6.4~keV line is produced by a transient photoionization source that was in a 
long-lasting ($>8.5$~years) bright state at $L_X \sim 10^{36}$~erg~s$^{-1}$ before a space telescope was able to detect 
it. No such source was detected with the {\it Einstein} observatory in 1979 (Watson et al. \cite{wat81}) and with 
subsequent X-ray observatories as well, which imposes a minimum distance of $\sim 4.6$~pc between the cloud emitting 
at 6.4~keV and the putative transient X-ray source. This distance is increased to $\sim 9.2$~pc if the cloud and the 
source are assumed to be at the same line-of-sight distance from the Earth. Furthermore, except for the extraordinarily 
long outburst of GRS~1915+105, which is predicted to last at least $\sim 20 \pm 5$~yr (Deegan et al. \cite{dee09}), the 
outburst duration of transient X-ray sources is generally much shorter than 8.5~years (see Degenaar et al. \cite{deg12}). 
We also note that the Arches cluster is probably too young ($t \sim 2.5$~yr; Najarro et al. \cite{naj04}) for an X-ray 
binary system to have formed within it.

Thus, the 6.4~keV line emission arising from the vicinity of the Arches cluster is unlikely to result from 
photoionization and is most probably produced by CR impact. We have shown that the measured slope of the nonthermal 
power-law continuum ($\Gamma = 1.6_{-0.2}^{+0.3}$) and the EW of the 6.4~keV line from this region 
(EW$_{\rm 6.4~keV}=1.2 \pm 0.2$~keV) are consistent with the predictions of the LECR ion model. On the other hand, LECR 
electrons cannot satisfactorily account for this emission, because it would require too high metallicity of the ambient 
gas ($Z > 3.1 \,Z_\odot$) and too low minimum energy $E_{\rm min} < 41$~keV (Table~\ref{Table4}). It is indeed unlikely 
that quasi-thermal electrons of such low energies can escape their acceleration region and penetrate a neutral or 
weakly ionized medium to produce the 6.4~keV line. We thus conclude that the 6.4~keV line emission from the cloud 
region is most likely produced by LECR ions.

\subsubsection{Are there other processes of production of the 6.4~keV line at work in the Arches cluster region?}

A relatively weak line at 6.4~keV is also detected in the spectrum of the X-ray emission from the star cluster.
The low EW of this line (EW$_{\rm 6.4~keV}=0.4\pm 0.1$~keV) may suggest that this radiation is 
produced by LECR electrons accelerated within the cluster. The existence of a fast electron population there is 
supported by the detection with the VLA of diffuse nonthermal radio continuum emission (Yusef-Zadeh et al. \cite{yus03}). 
The nonthermal electrons are thought to be produced by diffuse shock acceleration in colliding wind shocks of the 
cluster flow. 

It is, however, more likely that the 6.4~keV line detected from this region is produced in 
molecular gas along the line of sight outside the star cluster. In the {\it Chandra}/ACIS 6.4 keV line image of the Arches 
cluster region, the bow shock-like structure observed in the neutral or low-ionization Fe K$\alpha$ line covers the position 
of the star cluster (Wang et al. \cite{wan06}). The 6.4 keV line emission from the region ``Cluster" is not observed in the 
present map (Figs.~\ref{Fig15} and \ref{Fig19}), because it has been artificially removed in the process of 
subtraction of the continuum under the line (see Sect.~5.2). Lang et al. (\cite{lan02}) find evidence of molecular gas lying 
just in front of the ionized gas associated with the most eastern Arched filament (E1) close to the cluster sight line. According 
to the geometric arrangement of the $-30$~km~s$^{-1}$ clouds proposed by these authors, it is likely that the cluster is 
presently interacting with this foreground molecular gas. From the gradient of visual extinction detected by Stolte et al. 
(\cite{sto02}) over a field of $40'' \times 40''$ around the star cluster, $9 < \Delta A_V < 15$~mag, the H column density of 
this cloud along the line of sight can be estimated as $N_{\rm H}^C \gsim 3 \times 10^{22}$~cm$^{-2}$. The calculations 
of the present paper show that LECR ions can produce a significant 6.4 keV Fe K$\alpha$ line emission in such a cloud 
(Sect.~2), especially in the case of strong particle diffusion for which the CR path length $\Lambda$ can be much higher 
than $N_{\rm H}^C$ (see Appendix~A). It is thus likely that the weak nonthermal X-ray emission detected in the cluster 
spectrum has the same physical origin as the nonthermal emission from the cloud region. 

Relatively faint, diffuse emission in the neutral or low-ionization Fe K$\alpha$ line is also detected to the west 
of the Arches cluster, from an extended region centered at $(\ell,b)\sim (0.1^\circ,0.016^\circ$) (see Figs.~\ref{Fig19}). 
Capelli et al. (\cite{cap11b}) found the light curve of the 6.4~keV line flux from this region (labelled ''SN'' 
by these authors) to be constant over the 8-year observation time. With a measured proper motion of $\sim 4.5$~mas~yr$^{-1}$ 
almost parallel to the Galactic plane and towards increasing longitude (Stolte et al. \cite{sto08}; 
Clarkson et al. \cite{cla12}), the Arches cluster was located within this region of the sky $\sim 2 \times 10^4$~years ago. 
It is therefore conceivable that this emission is also due to LECR ions that were accelerated within or close to the 
cluster at that time. That the nonthermal X-ray emission is still visible today would then indicate that the fast 
ions have propagated since then in a medium of mean density $n_{\rm H} \lsim 10^3$~cm$^{-3}$. Indeed most of  the 6.4~keV line 
emission from LECR ions is produced by protons of kinetic energies $<$~200~MeV (see Fig.~\ref{Fig8}b) and the slowing-down 
time of 200-MeV protons in $n_{\rm H} = 10^3$~cm$^{-3}$ is $\sim 2 \times 10^4$~years.  

Capelli et al. (\cite{cap11b}) also considered a large region of 6.4~keV line emission located 
at $(\ell,b) \sim (0.11^\circ,0.075^\circ$) (see Fig.~\ref{Fig15}, {\it left panel}). They measured a fast 
variability of the neutral or low-ionization Fe K$\alpha$ line from this region and suggest that it could result 
from the illumination of a molecular cloud by a nearby transient X-ray source. The X-ray emission from this region 
is not studied in the present paper. 

\section{Origin of the LECR ion population}

Two sites of particle acceleration in the Arches cluster region have been proposed. As already 
mentioned, Yusef-Zadeh et al. (\cite{yus03}) report evidence of diffuse nonthermal radio synchrotron emission from 
the cluster and suggest that the emitting relativistic electrons are accelerated by diffuse shock acceleration in 
the colliding stellar winds of the cluster flow. Another scenario is proposed by Wang et al. (\cite{wan06}), who 
suggest that the 6.4~keV line emission from this region comes from LECR electrons produced in a bow shock resulting 
from an ongoing supersonic collision between the star cluster and an adjacent molecular cloud. Both processes could 
also produce LECR ions. 

Since the work of Wang et al. (\cite{wan06}), the apparent proper motion of the Arches cluster in the plane of 
the sky has been observed with Keck laser-guide star adaptive optics (Stolte et al. \cite{sto08}; Clarkson et al. 
\cite{cla12}). The direction of motion of the cluster stars relative to the field population is represented by the arrow 
in Fig.~\ref{Fig19}. The 6.4~keV line emission close to the cluster shows two bright knots connected by a 
faint bridge to the east of the cluster, i.e. ahead of the moving stars (Fig.~\ref{Fig19}). The overall structure 
indeed suggests a bow shock. However, the Fe K line intensity scales as the product of the density of cosmic rays 
and that of the ambient medium around the cluster, which is probably highly inhomogeneous. A clear bow shock shape is 
therefore not to be expected. In fact, the 6.4~keV map from this region may also be explained by LECR ions escaping from 
the cluster and interacting with adjacent molecular gas. Thus, the morphology of the bright structure at 6.4~keV does not 
allow us to favor one of the two proposed sites for the production of fast ions. But more information can be obtained by 
studying the CR power required to explain the X-ray emission (Sect.~7.1), as well as the accelerated particle composition 
(Sect.~7.2).

\subsection{CR spectrum and energetics}

Whether the main source of LECR ions in the Arches cluster region is the cluster bow shock or colliding stellar
 winds within the cluster flow, the nonthermal particles are likely to be produced by the diffusive shock acceleration 
(DSA) process. The nonthermal particle energy distribution resulting from this process can be written for linear 
acceleration as (e.g. Jones \& Ellison \cite{jon91})
\begin{equation} 
{dQ_{\rm DSA} \over dt} (E) \propto {p^{-s_{\rm DSA}} \over v}~,
\label{eq14}
\end{equation} 
where $p$ and $v$ are the particle momentum and velocity, respectively, and
\begin{equation} 
s_{\rm DSA} = {3\gamma_g -1 +4 M_S^{-2} \over 2 -2 M_S^{-2}}~.
\label{eq15}
\end{equation} 
Here, $\gamma_g$ is the adiabatic index of the thermal gas upstream the shock front ($\gamma_g=5/3$ for an 
ideal nonrelativistic gas) and $M_S=V_s/c_S$ is the upstream sonic Mach number of the shock, whose velocity 
is $V_s$. The sound velocity in the upstream gas is 
\begin{equation} 
c_S = \bigg({\gamma_g k T \over \mu m_{\rm H}}\bigg)^{1/2}~,
\label{eq16}
\end{equation} 
where $k$ is the Boltzmann constant, $T$ the gas temperature and $\mu m_{\rm H}$ the mean particle mass. 
In an interstellar molecular gas of temperature $T=100$~K, $c_S \approx 0.8$~km~s$^{-1}$. For a strong 
shock verifying $V_s \gg c_S$, we find from Eq.~(\ref{eq15}) $s_{\rm DSA} \cong 2$, such that the particle
spectrum in the nonrelativistic domain is a power law in kinetic energy of index $s\cong 1.5$ (see 
Eq.~(\ref{eq14})). Nonlinear effects due to the modification of the shock structure induced by the back-reaction 
of accelerated ions can slightly steepen the LECR spectrum, such that typically $1.5 < s < 2$ (Berezhko \& 
Ellison \cite{ber99}). The slope of the CR source spectrum that we derived from the X-ray spectral analysis, 
$s=1.9_{-0.6}^{+0.5}$ (see Table~\ref{Table4}), is consistent with this theory. 

The total power acquired by LECR ions in the cloud region can be estimated from the best-fit 
normalization of the nonthermal X-ray component ($N_{\rm LECR}$, see Table~\ref{Table4}). We find that the 
power injected by fast primary protons of energies between $E_{\rm min}=10$~MeV and 
$E_{\rm max}=1$~GeV in the X-ray emitting region is $(4.3^{+0.5}_{-0.2}) \times 10^{38}$~erg~s$^{-1}$ (still 
assuming a distance to the GC of 8~kpc). Taking the uncertainty in $E_{\rm min}$ into account 
changes the proton power to (0.2--1)$\times 10^{39}$~erg~s$^{-1}$ (see Sect.5.5). By integration of the 
CR source spectrum, we find that about 30--60\% more power is contained in suprathermal protons with 
$E < E_{\rm min}$, which, by assumption, do not penetrate dense regions of nonthermal X-ray production. 
Considering the accelerated $\alpha$-particles with $C_\alpha / C_p \cong 0.1$ adds another factor 
of 40\%. The required total CR power finally amounts to (0.5--1.8)$\times 10^{39}$~erg~s$^{-1}$. 

\subsubsection{Mechanical power available from massive star winds}

The total mechanical power contained in the fast winds from massive stars of the cluster can be 
estimated from near infrared and radio data. Using such observations, Rockefeller et al. (\cite{roc05}) 
modeled the diffuse thermal X-ray emission from the cluster with 42 stellar wind sources with mass-loss rates
in the range $(0.3$--$17)\times 10^{-5}~M_\odot$~yr$^{-1}$ and a terminal wind velocity of 1000~km~s$^{-1}$. 
The total mechanical power contained in these 42 sources is $4 \times 10^{38}$~erg~s$^{-1}$. 
Of course, only a fraction of this energy reservoir can be converted to CR kinetic energy. We also note 
that LECR ions produced in the cluster are likely to diffuse away isotropically, such that 
those interacting with an adjacent molecular cloud emitting at 6.4 keV would probably represent a 
minority. Thus, the cluster wind is likely not powerful enough to explain the intensity of the nonthermal
X-ray emission. 

\subsubsection{Mechanical power available from the Arches cluster proper motion}

The proper motion of the Arches cluster relative to the field star population has recently been measured to be 
$172 \pm 15$~km~s$^{-1}$ (Stolte et al. \cite{sto08}; Clarkson et al. \cite{cla12}).The cluster is also moving 
away from the Sun, with a heliocentric line-of-sight velocity of $+95 \pm 8$~km~s$^{-1}$ (Figer et al. 
\cite{fig02}). The resulting three-dimensional space velocity is $V_* \approx 196$~km~s$^{-1}$. To model 
the form of the bow shock resulting from this supersonic motion, we approximate the cluster as a point source 
object that loses mass at a rate $\dot{M}_W=10^{-3}~M_\odot$~yr$^{-1}$ through a wind of terminal velocity 
$V_W=1000$~km~s$^{-1}$ (see Rockefeller et al. \cite{roc05}). The shape of the bow shock is determined by the 
balance between the ram pressure of the cluster wind and the ram pressure of the ongoing ISM gas. The 
pressure equilibrium is reached in the cluster direction of motion at the so-called standoff distance from 
the cluster (see, e.g., Wilkin \cite{wil96})
\begin{equation}
R_{\rm bs} =  \bigg({\dot{M}_W V_{W}\over 4 \pi \rho_{\rm IC} V_*^2}\bigg)^{0.5} = 2.4~{\rm pc}~, 
\label{eq17}
\end{equation}
where $\rho_{\rm IC} \cong 1.4 m_p n_{\rm IC}$ is the mass density of the local ISM. Here, we assume that since 
the birth of the cluster $\sim$2.5~Myr ago (Figer et al. \cite{fig02}; Najarro et al. \cite{naj04}), the bow 
shock has propagated most of the time in an intercloud medium of mean H density $n_{\rm IC} \sim 10$~cm$^{-3}$ (see 
Launhardt et al. \cite{lau02} for a description of the large-scale ISM in the GC region). 

The circular area of a bow shock projected on a plane perpendicular to the direction of motion is 
$A_{\rm bs} \sim 10 \pi R_{\rm bs}^2$ (see Wilkin \cite{wil96}). Thus, the mechanical power processed by 
the cluster bow shock while propagating in the intercloud medium is 
$P_{\rm IC}=0.5\rho_{\rm IC}V_*^3A_{\rm bs}\sim 1.5\times 10^{38}$~erg~s$^{-1}$. In comparison, the
steady state, mechanical power supplied by supernovae in the inner $\sim$200~pc of the Galaxy is 
$\sim 1.3\times 10^{40}$~erg~s$^{-1}$ (Crocker et al. \cite{cro11}). LECRs continuously accelerated out of 
the intercloud medium at the Arches cluster bow shock possibly contribute $\sim$1\% of the steady-state 
CR power in the GC region (assuming the same acceleration efficiency as in supernova remnants). 

The initial total kinetic energy of the cluster motion is $0.5M_*V_*^2\sim 1.9\times 10^{52}$~erg, where 
$M_* \sim 5 \times 10^4~M_\odot$ is the cluster initial total mass (Harfst et al. \cite{har10}). This 
energy would be dissipated in $\sim$4~Myr according to our estimate of $P_{\rm IC}$. 

Most of the interstellar gas mass in the Galactic nuclear bulge is contained in dense molecular 
clouds with average H densities of $n_{\rm MC} \sim 10^4$~cm$^{-3}$ and a volume filling factor of a few 
percent (Launhardt et al. \cite{lau02}). In the region where the Arches cluster is presently located, the 
volume filling factor of dense molecular gas is even $\gsim$0.3 (Serabyn \& G\"uesten \cite{ser87}). Thus, 
the probability of a collision between the cluster bow shock and a molecular cloud is strong. The 
evidence that the cluster is presently interacting with a molecular cloud has already been discussed by Figer 
et al. (\cite{fig02}) and Wang et al. (\cite{wan06}). This molecular cloud was identified as ``Peak 2" in 
the CS map of Serabyn \& G\"uesten (\cite{ser87}), who estimated its mass to be 
$M_{\rm MC}=(6\pm 3)\times 10^4~M_\odot$ and mean H density as $n_{\rm MC}=(2\pm 1)\times 10^4$~cm$^{-3}$. 
The corresponding diameter for a spherical cloud is $d_{\rm MC} \sim 5.5$~pc or 2.4' at a 
distance of 8~kpc, which is consistent with the apparent size of the cloud (Serabyn \& G\"uesten 
\cite{ser87}). 

The total kinetic power processed in this collision is given by 
\begin{equation}
P_{\rm MC}={1 \over 2} \rho_{\rm MC}(V_*+V_{\rm MC})^3 A_C~, 
\label{eq18}
\end{equation}
where $\rho_{\rm MC} \cong 1.4 m_p n_{\rm MC}$, $V_{\rm MC}$ is the velocity of the  
molecular cloud projected onto the direction of motion of the Arches cluster, and $A_C$ the area of the 
contact surface between the ``Peak 2" cloud and the bow shock. The latter quantity is not well known. We 
assume that it is equal to the area of the large region around the cluster emitting in the 6.4~keV line 
(i.e., the region labeled ``Cloud" in Fig.~\ref{Fig15} and Table~\ref{Table2}): $A_C=7$~pc$^2$. 

The cloud-projected velocity $V_{\rm MC}$ obviously depends on the orbital path of the molecular cloud about 
the GC. By studying the velocity field of the molecular gas around the Arches cluster, Lang et al. 
(\cite{lan01,lan02}) obtained constraints on the trajectory of the $-30$~km~s$^{-1}$ clouds. They 
find that the cloud complex likely resides on the far side of the GC, either on a $x_2$ orbit 
(a noncircular orbit family set up in response to the Galaxy's stellar bar) or on a trajectory directed towards 
Sgr~A$^\ast$ (the cloud complex would then be radially infalling into the supermassive black hole) or perhaps on 
a trajectory midway between the two situations. If the ``Peak 2" cloud resides on an $x_2$ orbit in the back
of the Galaxy, the collision of this cloud with the Arches cluster is almost frontal (see Stolte et al. 
\cite{sto08}; Clarkson et al. \cite{cla12}), and $V_{\rm MC}$ is close to the $x_2$ orbital speed 
$v_{\rm orb} \sim 80$~km~s$^{-1}$ (see, e.g., Molinari et al. \cite{mol11}). But if the cloud is radially infalling 
towards the supermassive black hole, given the radial velocity of the cloud ($v_{\rm rad}\approx-30$~km~s$^{-1}$) 
and the radial and transverse velocity components of the Arches cluster ($v_{\rm rad}\approx+95$~km~s$^{-1}$ and 
$v_{\rm trans}\approx+172$~km~s$^{-1}$ directed towards positive longitude; see Clarkson et al. \cite{cla12}), one 
finds that $V_{\rm MC} \sim 20$~km~s$^{-1}$. Thus, depending on the exact cloud trajectory 
$V_{\rm MC} \approx 50 \pm 30$~km~s$^{-1}$, which, together with $V_* \approx 196$~km~s$^{-1}$, gives 
$P_{\rm MC} \sim 2.3 \times 10^{40}$~erg~s$^{-1}$ from Eq.~(\ref{eq18}). 

In comparison, the CR power needed to explain the X-ray observations is $dW/dt=(0.5$--1.8)$\times 10^{39}$~erg~s$^{-1}$, 
such that the required particle acceleration efficiency in the bow shock system amounts to a few percent. 
This is a typical efficiency in the DSA theory and in the phenomenology of the acceleration of the Galactic CRs 
in supernova remnant shocks, as well (see, e.g., Tatischeff \cite{tat08} and references therein). However, a detailed 
study of the particle acceleration process at work in this peculiar shock system would go beyond the scope of this paper.  

The collision kinetic power estimated above is comparable to the steady-state, mechanical power 
due to supernovae in the inner Galaxy, $\sim 1.3\times 10^{40}$~erg~s$^{-1}$ (Crocker et al. \cite{cro11}). 
But the typical time duration of a collision between the Arches cluster bow shock and a molecular cloud 
is expected to be only $\sim 3\times 10^4$~yr, assuming that the size $d_{\rm MC} \sim 5.5$~pc is typical 
of the highly-fragmented dense molecular gas of the GC region. The mechanical energy released 
in such a collision is then $\sim$10$^{52}$~erg, i.e. comparable to the cluster initial total kinetic energy. 
Thus, the cluster bow shock very likely has collided no more than once with a molecular cloud since the cluster 
birth $\sim$2.5~Myr ago. Such a collision can briefly release in the ISM a power in LECRs comparable to 
the steady-state CR power supplied by supernovae in the GC region (Crocker et al. \cite{cro11}).

The ISM volume swept up by the Arches cluster bow shock since the cluster birth can be estimated as
\begin{equation}
V_{\rm bs} = A_{\rm bs} V_* t_* \sim 9 \times 10^4~{\rm pc^3}~,
\label{eq18p}
\end{equation}
where $t_*=2.5$~Myr is the estimated cluster age. This volume represents less than 1\% of the total volume 
of the Galactic nuclear bulge, $V_{\rm NB} \sim 1.5 \times 10^7$pc$^3$. In comparison, the  volume filling factor 
of dense molecular cloud in the inner $\sim$230~pc of the Galaxy is a few percent (Launhardt et 
al. \cite{lau02}). It is thus likely that the star cluster did not experience any interaction with a molecular 
cloud before the one with the ``Peak 2" cloud that is presently observed. 

Simulated orbits of the Arches cluster about the GC suggest that the cluster formed in the front of the Galaxy 
near an $x_2$ orbit (Stolte et al. \cite{sto08}). That the Arches cluster presently interacts 
with a molecular cloud located behind the GC shows that the cluster's orbit is retrograde to the general motion of 
stars and gas clouds in the bar potential (see Fig.~8 of Stolte et al. \cite{sto08}). According to the simulation of 
possible orbits, the cluster has performed about half a revolution around the GC since its formation, which may have 
brought it near the far side of the elliptical ring of dense molecular clouds recently studied with the {\it Herschel} 
satellite (Molinari et al. \cite{mol11}). In this environment, the probability of a collision between the cluster and 
a molecular cloud has become strong.

\subsection{Accelerated ion composition}

Fast C and heavier ions can emit very broad X-ray lines resulting from $2p$ to $1s$ (K$\alpha$) and 
$3p$ to $1s$  (K$\beta$) in-flight transitions. The $2p$ and $3p$ orbital states can be populated 
either by electron capture from ambient atoms (i.e. charge exchange) or by excitation of $1s$ 
electrons for fast ions having one or two electrons. To study the composition of the energetic ions 
accelerated near the Arches cluster, we developed new LECR ion models that include the line emission of 
fast C, N, O, Ne, Mg, Si, S, and Fe. We used the tables of K X-ray differential multiplicities, $dM_i^{Kk}/dE$, 
given in Tatischeff et al. (\cite{tat98}). This quantity is defined as the 
number of photons emitted in the $Kk$ line by the projectile $i$ as it slows down over the 
differential kinetic energy interval $dE$, owing to interactions with all the constituents of the 
ambient medium. In the adopted steady-state, slab interaction model, the X-ray line production rate 
is then simply given by
\begin{equation}
{dQ_i^{Kk} \over dt}(E_X) =  \int_0^\infty {dM_i^{Kk} \over dE}(E_X,E) dE \int_{E}^{E_\Lambda^i(E)} 
{dQ_i \over dt} (E') dE'~.
\label{eq19}
\end{equation}
The energy of the emitted X-rays depends on both the velocity and spatial distributions of the fast 
ions through the usual Doppler formula. We assumed isotropic propagation of the LECRs in the 
interaction region, which leads to a maximum broadening of the lines. 

We note that the calculations of Tatischeff et al. (\cite{tat98}) were done for an ambient medium of 
solar composition. Nevertheless, their multiplicity results can be used in good approximation for a 
medium of order twice solar metallicity, since it was found that most of the line emission from the fast 
ions is produced by interactions with ambient H and He. 

We first developed an X-ray production model in which the abundances of the heavy ions are in solar 
proportions relative to each other, but can vary with respect to H and He, that is (see also Eq.~(\ref{eq7}))
\begin{equation}
{C_i \over C_p} = f_{\rm met} \bigg({C_i \over C_p}\bigg)_\odot~. 
\label{eq20}
\end{equation}
The parameter $f_{\rm met}$ adds to the spectral index $s$ and the metallicity of the ambient medium 
$Z$ as a free parameter of the model. But here, we fixed $E_{\rm min}=10$~MeV and $\Lambda=5 \times 10^{24}$~cm$^{-2}$, 
as in Sect.~5.2. The resulting model was made readable in XSPEC and then used to fit the {\it XMM-Newton} 
spectra extracted from the cloud region. As before, we chose to fit to the data the global XSPEC model 
WABS$\times$(APEC + LECR$i$). The best fit was obtained for $f_{\rm met}=3.9_{-3.9}^{+11.3}$. The corresponding 
$\chi^2$ and best-fit values of the other parameters are nearly the same as in Table~\ref{Table4}, in 
particular with $Z/Z_\odot=1.7\pm0.2$ and $s=2.0_{-0.7}^{+0.8}$. The obtained limit at the 90\% confidence level 
$f_{\rm met}<15.2$ shows that the heavy ion abundance is not well constrained by this model. 

   \begin{figure}
   \centering
   \includegraphics[width=7.5cm]{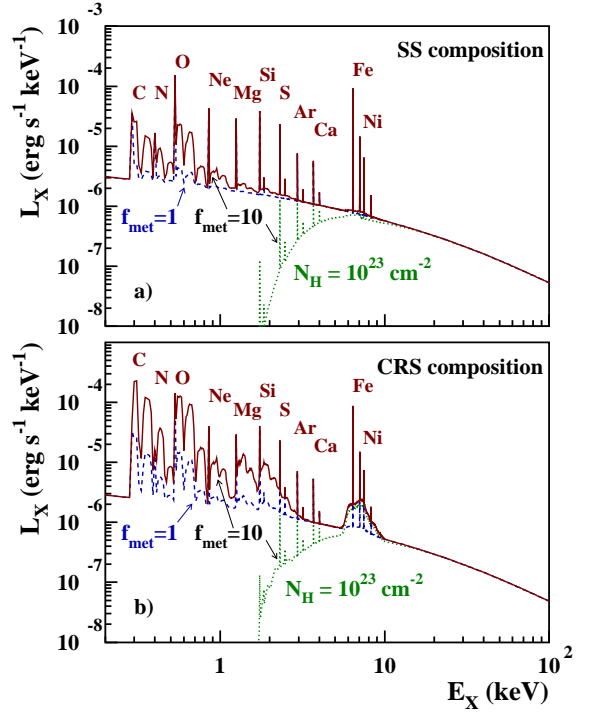}
      \caption{Calculated X-ray emission produced by LECR ions with the spectral parameters $s=1.9$, 
      $E_{\rm min}=10$~MeV~nucleon$^{-1}$, and the escape path length $\Lambda=5\times 10^{24}$~cm$^{-2}$,
      interacting in a gas cloud of metallicity $Z=1.7\,Z_\odot$ (as obtained in Sect.~5). In panel {\bf a)} the 
      abundances of the accelerated heavy ions are in solar proportion, but with a metallicity enhancement 
      factor relative to the solar system (SS) composition (see text) $f_{\rm met}=1$ ({\it dashed curve}) 
      and $f_{\rm met}=10$ ({\it solid} and {\it dotted curves}). Panel {\bf b)}: same but for the CRS 
      composition (see text). The {\it dotted curves} show the effect of photoelectric absorption on the 
      X-ray spectra for a H column density of 10$^{23}$~cm$^{-2}$. 
              }
         \label{Fig20}
   \end{figure}

Figure~\ref{Fig20}a shows calculated X-ray spectra for the solar system composition with 
$f_{\rm met}=1$ and 10. We see that intense broad lines can be emitted below 1 keV from 
X-ray transitions in fast C, N, and O. But this line emission cannot be observed from sources in the 
GC region, because of the strong interstellar photoelectric absorption along the line of 
sight. With $N_{\rm H} \approx 10^{23}$~H~cm$^{-2}$ (Table~\ref{Table4}), the main constraint on the 
accelerated particle composition is provided by a very broad line feature between $\sim$5.5 and 9~keV 
owing to de-excitations in fast Fe. This emission is produced by Fe ions of energies between $\sim$5 
and 20~MeV~nucleon$^{-1}$ (Tatischeff et al. \cite{tat98}). 

In a second model, we assumed that the fast metals have the composition as the current epoch Galactic 
CRs at their sources. We obtained the CR source (CRS) composition by taking the heavy ion abundances 
relative to O given by Engelmann et al. (\cite{eng90}) and using the abundance ratios 
$C_\alpha/C_{\rm O}=19$ and $C_p/C_\alpha=15$ recommended by Meyer et al. (\cite{mey97}). The resulting 
CRS composition is consistent with the recent theoretical works of Putze et al. (\cite{put11}). The Fe 
abundance in the CRS composition is $C_{\rm Fe}/C_p=6.72 \times 10^{-4}$, which is 19 times higher than 
the one in the solar system composition, $(C_{\rm Fe}/C_p)_\odot=3.45 \times 10^{-5}$ (Lodders et al. 
\cite{lod03}). The best fit of this model to the X-ray spectra of the cloud region was obtained 
for $f_{\rm met}=0.11_{-0.11}^{+0.60}$ (i.e. $f_{\rm met}<0.71$ at the 90\% confidence level), consistent 
with the higher abundance of Fe in the CRS composition. X-ray spectra calculated for this composition 
are shown in Fig.~\ref{Fig20}b. 

The main outputs of this analysis are the nondetection in the X-ray spectra of the cloud region of a 
significant excess emission from fast Fe and the implication that $C_{\rm Fe}/C_p \lsim 5 \times 10^{-4}$. 
This result by itself does not provide strong support for one or the other possible site of acceleration 
of the LECR ions in the Arches cluster region. Indeed, with the best-fit metallicity $Z / Z_\odot = 1.7 \pm 0.2$, 
the Fe abundance in the local molecular cloud is
$a_{\rm Fe} = Z/Z_\odot \, (a_{\rm Fe})_\odot \approx 5.9 \times 10^{-5}$, which is well below the upper 
limit obtained above. The Fe abundance can be slightly higher in the wind material expelled by the massive 
stars of the Arches cluster. According to Parizot et al. (\cite{par97}), one expects in the average 
composition of the winds from OB associations in the inner Galaxy $a_{\rm Fe} = 1.4 \times 10^{-4}$, which is 
still below the derived upper limit. 

However, the Galactic CRS composition is best described in terms of a general enhancement of the 
refractory elements such as Fe relative to the volatile ones (Meyer et al. \cite{mey97}). This selection 
effect is most likely related to the acceleration process at work in supernova remnant shock waves. Given 
the limit $f_{\rm met}<0.71$ that we obtained for the CRS composition, we conclude that this effect is 
weaker in the shock system associated with the Arches cluster proper motion, which according to the 
energetics arguments discussed in Sect.~6.2.2, is the most likely site of acceleration of the X-ray 
emitting LECR ions. 

\section{Cosmic-ray ionization rate}

We estimated in Sect.~7.1 that a kinetic power of (0.5--1.8)$\times 10^{39}$~erg~s$^{-1}$ is currently 
delivered by the Arches cluster bow shock system to LECR ions of energies $< 1$~GeV~nucleon$^{-1}$. The power 
continuously deposited into the adjacent molecular cloud is lower than that, because of (i) 
the nonpenetration of CRs with $E < E_{\rm min}$ into the interaction region and (ii) the escape from 
the cloud of the highest energy CRs. For $\Lambda=5 \times 10^{24}$~cm$^{-2}$, protons of energies up to 
180~MeV are stopped in the cloud, whereas those injected at higher energies do not virtually lose energy 
in this medium. Taking these effects into account, we find that the power deposited by 
LECRs into the cloud amounts to $\dot{W}_d \sim 4 \times 10^{38}$~erg~s$^{-1}$. The initial kinetic 
energy of the fast ions is essentially lost through ionization of the ambient gas and the corresponding 
ionization rate can be estimated to be
\begin{equation}
\zeta_{\rm H} = {1.4 m_p \dot{W}_d \over \epsilon_i M_{\rm MC}} \sim 10^{-13} {\rm~H^{-1}~s^{-1}}~, 
\label{eq21}
\end{equation}
where $\epsilon_i \approx 40$~eV is the mean energy required for a fast ion to produce a free electron 
in a neutral gas mixture of H$_2$ and He in cosmic proportion (Dalgarno et al. \cite{dal99}) and 
$M_{\rm MC}=(6\pm 3)\times 10^4~M_\odot$ is the cloud mass (Serabyn \& G\"uesten \cite{ser87}). The 
mean ionization rate induced by LECRs in this molecular cloud is significantly higher than the mean 
ionization rate in the GC region, $\zeta_{\rm H}\gsim 10^{-15}$~H$^{-1}$~s$^{-1}$ (see 
Crocker et al. \cite{cro11} and references therein). 

By integrating the differential equilibrium number of LECRs in the X-ray production region (see 
Eq.~(\ref{eq1})), we find that the total kinetic energy contained in fast ions diffusing in the cloud 
is $E_{\rm tot} \sim 4 \times 10^{48}$~erg. The corresponding mean energy density is 
$E_{\rm tot}/V_{\rm MC} \sim 1000$~eV~cm$^{-3}$ (here $V_{\rm MC}=M_{\rm MC}/(1.4n_{\rm MC}m_p)$), which is  
about one thousand times higher than the Galactic CR energy density in the solar neighborhood. Thus, 
the molecular cloud irradiated by fast particles accelerated near the Arches cluster bow shock shows some
similarities with the ``extreme CR dominated regions" recently studied by Papadopoulos et al. 
(\cite{pap11}) in the context of starbursts. Following the works of these authors, LECRs could 
explain the high temperature of the ``Peak 2" cloud measured by Serabyn \& G\"uesten (\cite{ser87}): 
$T \gsim 100$~K. 

\section{Gamma-ray counterparts}

Collisions of LECR ions with molecular cloud matter can lead to nuclear excitations of both ambient 
and accelerated heavy ions followed by emission of de-excitation $\gamma$-ray lines (Ramaty et al. 
\cite{ram79}; Benhabiles-Mezhoud et al. \cite{ben11}). We calculated the $\gamma$-ray line flux 
expected from the Arches cluster region using the same CR interaction model as before (see Appendix~A), 
with $s=1.9$, $E_{\rm min}=10$~MeV, $N_{\rm LECR}=5.6\times 10^{-8}$~erg~cm$^{-2}$~s$^{-1}$, 
$\Lambda=5\times 10^{24}$~cm$^{-2}$, and $Z/Z_\odot = 1.7$ (Table~\ref{Table4}). 
The predicted flux is well below the sensitivity limit of the {\it INTEGRAL} observatory. For example, 
we obtain a flux of $2.3 \times 10^{-8}$~ph~cm$^{-2}$~s$^{-1}$ in the 4.44~MeV line from de-excitations 
of ambient $^{12}$C, whereas the sensitivity of the {\it INTEGRAL} spectrometer SPI for detection of a 
narrow line at this energy is $> 10^{-5}$~ph~cm$^{-2}$~s$^{-1}$.

Nuclear interactions of CR ions with ambient matter can also lead to high-energy $\gamma$-ray emission 
via the production and subsequent decay of $\pi^0$ mesons. We used the model of Dermer (\cite{der86}) 
for this calculation, but multiplied the $\pi^0$ emissivity given by this author by a factor of 1.27 to 
be consistent with the local emissivity measured with the {\it Fermi} Gamma-ray Space Telescope (Abdo et 
al. \cite{abd09}). The high-energy $\gamma$-ray flux strongly depends on the shape of the CR 
energy distribution, because the neutral pions are produced at significantly higher energies than the
nonthermal X-rays. We first assumed a CR source spectrum of the form given by Eq.~(\ref{eq14}) for this 
calculation (i.e. resulting from the DSA process) with $s_{\rm DSA}=2s-1=2.8$ and no high-energy 
cutoff. We then found that the Arches cluster region would emit a flux of 
$5.7 \times 10^{-7}$~ph~cm$^{-2}$~s$^{-1}$ in $\gamma$-rays of energies $> 300$~MeV. Such high-energy 
emission would have probably been already detected by {\it Fermi}, since the predicted flux is $\sim$1.75 times 
higher than the flux of the Galactic central source 1FGL J1745.6--2900 (Chernyakova et al. \cite{che11}). 
But an exponential cutoff in the CR distribution can be expected either because of the finite size of the 
particle acceleration region near the cluster bow shock or the finite time available for 
particle acceleration. For example, with an exponential cutoff at 0.5~GeV~nucleon$^{-1}$ 
(resp. 1~GeV~nucleon$^{-1}$), the flux of $\gamma$-rays $> 300$~MeV would be reduced to 
$1.4 \times 10^{-8}$~ph~cm$^{-2}$~s$^{-1}$ (resp. $5.1 \times 10^{-8}$~ph~cm$^{-2}$~s$^{-1}$) without 
significantly changing the nonthermal X-ray production. The high-energy $\gamma$-ray emission from the 
Arches cluster region would then be undetectable with {\it Fermi}. 

\section{Summary}

We have studied the production of nonthermal line and continuum X-rays by interaction of 
LECR electrons and ions with a neutral ambient medium in detail. We developed a steady-state, slab model in 
which accelerated particles penetrate at a constant rate a cloud of neutral gas, where they 
produce nonthermal X-rays by atomic collisions until they either stop or escape from the cloud. We 
examined the properties of the neutral Fe K$\alpha$ line excited by impacts of LECR electrons 
and ions. The predicted line EW and luminosity, as well as the slope of the underlying 
bremsstrahlung continuum, were presented as functions of the free parameters of the model. These 
results are intended to help observers study the potential role of LECRs for any 6.4~keV line 
emission and possibly decipher the nature of the nonthermal particles responsible for the line
emission. In addition, we generated LECR electron and ion models that can be used in the XSPEC 
software for more quantitative comparison with data. 

We showed, in particular, that the EW of the neutral Fe \kalpha line excited by LECR electrons is 
generally expected to be lower than 1~keV, except if the metallicity of the ambient medium 
exceeds $\approx$2$\,Z_\odot$. But LECR ions with a relatively soft source spectrum can lead to a 
much larger EW. However, the production of 6.4~keV line photons by both LECR electrons and ions is 
relatively inefficient: the radiation yield $R_{6.4~{\rm keV}}=L_X(6.4~{\rm keV}) / (dW/dt)$ is 
typically on the order of 10$^{-6}$, meaning that a high power in LECRs should generally be needed 
to produce an observable neutral Fe K$\alpha$ line. 

We then employed the newly developed models to study the X-ray emission emanating from the Arches 
cluster region. We used all public {\it XMM-Newton} EPIC observations encompassing the studied region 
for our analysis. The main results of this analysis can be summarized as follows. 
\begin{itemize}
\item The X-ray flux detected from the Arches cluster is dominated by the emission of an optically 
thin thermal plasma with a temperature $kT \sim 1.7$~keV. This component most likely arises from 
the thermalization of massive star winds that merge and expand together, plus the contribution of 
several colliding stellar wind binaries within the cluster. 
\item A second thermal plasma of lower temperature is required to explain the presence of He-like Si 
and S \kalpha emission lines in the X-ray spectrum of the cluster. This component, which was not 
detected in previous X-ray observations, may be produced by a collection of individual massive stars. 
\item Bright 6.4 keV Fe \kalpha line structures are observed around the Arches cluster. 
We found that the line flux from this region is consistent with its being constant over more than eight 
years of {\it XMM-Newton} repeated observations, in agreement with the recent works of Capelli et al. 
(\cite{cap11b}). This radiation is unlikely to result from the photoionization of a molecular cloud
by a hard X-ray source. It is also probably not produced by LECR electrons, because it would require a 
metallicity of the ambient gas ($Z > 3.1 \,Z_\odot$) that is too high. On the other hand, the X-ray emission 
observed around the cluster can be well-fitted with a model composed of an optically thin thermal 
plasma and a nonthermal component produced by LECR ions. The best-fit metallicity of the ambient medium 
found with this model is $Z / Z_\odot = 1.7 \pm 0.2$, and the best-fit CR source spectral index is 
$s=1.9_{-0.6}^{+0.5}$.
\item The required flux of LECR ions is likely to be produced by the diffusive shock acceleration process in 
the region of interaction of the Arches cluster and the adjacent molecular cloud identified as ``Peak 2'' in 
the CS map of Serabyn \& G\"uesten (\cite{ser87}). We estimated that a total kinetic power of 
$\sim 2.3 \times 10^{40}$~erg~s$^{-1}$ is currently processed in the ongoing supersonic collision between the 
star cluster and the molecular cloud emitting the 6.4 keV line. A particle acceleration efficiency of a 
few percent in the resulting bow shock system would produce enough CR power to explain the luminosity of the 
nonthermal X-ray emission. 
\item We developed LECR ion models that include the production of broad X-ray lines from fast C and 
heavier ions following electron captures from ambient atoms (i.e. charge exchanges) and atomic 
excitations. It allowed us to constrain the abundance of fast Fe ions relative to protons in the LECR 
ion population: $C_{\rm Fe}/C_p \lsim 5 \times 10^{-4}$. This limit is $\sim$15 times higher than the Fe 
abundance in the solar system composition. 
\item The mean ionization rate induced by LECRs in the molecular cloud that is thought to presently 
interact with the Arches cluster is $\zeta_{\rm H}\sim 10^{-13}$~H$^{-1}$~s$^{-1}$. The CR energy density 
in the interaction region is estimated to be $\sim 1000$~eV~cm$^{-3}$, which is about one thousand times 
higher than the Galactic CR energy density in the solar neighborhood. 
\item The high-energy $\gamma$-ray emission produced by hadronic collisions between CRs accelerated in 
the Arches cluster bow shock system and ambient material might be detected with the {\it Fermi} Gamma-ray 
Space Telescope. It crucially depends, however, on the unknown shape of the CR energy distribution above 
$\sim$1~GeV~nucleon$^{-1}$.
\end{itemize}

The nonthermal X-ray emission emanating from the Arches cluster region probably offers the best available 
signature currently for a source of low-energy hadronic cosmic rays in the Galaxy. Deeper observations of this 
region with X-ray telescopes would allow better characterization of the acceleration process 
and the effects of LECRs on the interstellar medium. The theory presented in this paper could also be 
useful for identifying new sources of LECRs in the Galaxy. 

\begin{acknowledgements}
We would like to thank an anonymous referee for suggestions that helped improve the paper. G. M. 
acknowledges financial support from the CNES. This work uses observations performed with {\it XMM-Newton}, an ESA 
Science Mission with instruments and contributions directly funded by ESA member states and the USA (NASA). 
\end{acknowledgements}

\begin{appendix}

\section{Cosmic-ray interaction model}

\begin{figure}
   \centering
   \includegraphics[width=7.5cm]{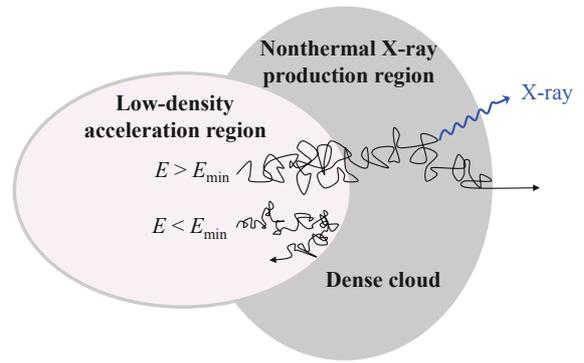}
      \caption{Schematic illustration of the cosmic-ray interaction model: fast particles
      produced in a low-density acceleration region can diffusively penetrate 
      a denser cloud (if their kinetic energy is higher than a threshold energy 
      $E_{\rm min}$) and then produce nonthermal X-rays by atomic collisions.}
         \label{Fig1}
   \end{figure}

We consider a model in which low-energy cosmic rays (LECRs) are produced in an unspecified
acceleration region and penetrate a nearby cloud of neutral gas at a constant rate (see 
Fig.~\ref{Fig1}). The energetic particles can produce nonthermal X-rays by atomic collisions while 
they slow down by ionization and radiative energy losses in the dense cloud. We further assume 
that the LECRs that penetrate the cloud can escape from it after an energy-independent 
path length $\Lambda$, which is a free parameter of the model. The 
differential equilibrium number of primary CRs of type $i$ (electrons, protons, or
$\alpha$ particles) in the cloud is then given by
\begin{equation} 
N_i (E) = {1 \over [dE/dt(E)]_i} \int_{E}^{E_\Lambda^i(E)} {dQ_i \over dt} (E') dE'~,
\label{eq1}
\end{equation} 
where $(dQ_i/dt)$ is the differential rate of LECRs injected in the cloud, 
$[dE/dt(E)]_i$ is the CR energy loss rate, and the maximum energy 
$E_\Lambda^i(E)$ is related to the escape path length $\Lambda$ (expressed in units of
H atoms cm$^{-2}$) by 
\begin{equation} 
\Lambda = \int_{E}^{E_\Lambda^i(E)} {dE' \over [dE/dN_{\rm H}(E')]_i}~,
\label{eq2}
\end{equation} 
where
\begin{equation} 
\bigg({dE \over dN_{\rm H}}\bigg)_i  = {1 \over v_i n_{\rm H}} \bigg({dE \over dt}\bigg)_i \simeq 
m_p \bigg[ \bigg({dE \over dx}\bigg)_{i,{\rm H}} + 4 \, a_{\rm He} \bigg({dE \over dx}\bigg)_{i,{\rm He}} \bigg]~. 
\label{eq3}
\end{equation} 
Here, $v_i$ is the particle velocity, $n_{\rm H}$ the mean number density of H atoms in the
cloud, $m_p$ the proton mass, $a_{\rm He}=0.0964$ the cosmic abundance of He relative 
to H (Lodders \cite{lod03}), and $(dE/dx)_{i,{\rm H}}$ and $(dE/dx)_{i,{\rm He}}$ the CR 
stopping powers (in units of MeV g$^{-1}$ cm$^{2}$) in ambient H and He, respectively. We 
used for electrons the stopping-power tables of Berger \& Seltzer (\cite{ber82}) below 
1~GeV and the relativistic formulae given by Schlickeiser (\cite{sch02}) above this energy. 
The stopping powers for protons and $\alpha$-particles were extracted from the online 
databases PSTAR and ASTAR, respectively (Berger et al. \cite{ber05}). 

The process of CR transport in the cloud, which does not need to be specified in the above 
formalism, is nevertheless relevant to estimate the escape path length $\Lambda$ from the 
cloud size. It is clear that if the cloud medium is not diffusive, because of, e.g., efficient 
ion-neutral damping of MHD waves, $\Lambda \sim n_{\rm H} L_C$, where $L_C$ is the characteristic 
size of the cloud. But otherwise, the escape path length, which can then be estimated as
\begin{equation} 
\Lambda \sim {L_C^2 v_i n_{\rm H} \over 6 D}~, 
\label{eq4}
\end{equation} 
can be much greater than the characteristic column density $N_{\rm H}^C = n_{\rm H} L_C$, depending 
on the diffusion coefficient $D$. For example, with the typical mean diffusion coefficient for the 
propagation of Galactic CR nuclei in the local interstellar magnetic field $B$ (Berezinsky et al. 
\cite{ber90}),
\begin{equation} 
D \approx 10^{28}\beta \bigg({R_i \over 1{\rm~GV}}\bigg)^{0.5} \bigg({B \over 3{\rm~\mu G}}\bigg)^{-0.5}
~~{\rm cm}^2~{\rm s}^{-1},
\label{eq5}
\end{equation} 
where $\beta=v_i/c$ and $R_i$ is the particle rigidity, one gets from Eq.~(\ref{eq4}) for non-relativistic protons: 
\begin{eqnarray} 
\Lambda & \sim & 5 \times 10^{24}\bigg({E \over 50{\rm~MeV}}\bigg)^{-0.25} \nonumber \\ 
& & \times~ \bigg({N_{\rm H}^C \over 10^{23}{\rm~cm^{-2}}}\bigg)^2 
\bigg({n_{\rm H} \over 10^{4}{\rm~cm^{-3}}}\bigg)^{-1} 
\bigg({B \over 100{\rm~\mu G}}\bigg)^{0.5}~~{\rm cm}^{-2}, 
\label{eq6}
\end{eqnarray} 
where $N_{\rm H}^C$, $n_{\rm H}$ and $B$ are scaled to typical values for massive molecular
clouds in the GC region. 

For nonrelativistic particles diffusing in the cloud with a diffusion coefficient 
$D \propto \beta R_i^{s_D}$ typically with $0.3 < s_D < 0.5$, the escape path length 
estimated from Eq.~(\ref{eq4}) depends only mildly on energy as $\Lambda \propto E^{-s_D/2}$. 
However, we have adopted here a simple slab model with an energy-independent 
escape path length in order to limit the number of free parameters as much as possible. 

The process of CR penetration into molecular clouds is not well known (see, e.g., Gabici et al. 
\cite{gab07} and references therein). The theoretical predictions range from almost-free penetration 
(e.g. Cesarsky \& V\"olk \cite{ces78}) to exclusion of CRs of kinetic energies up to tens of GeV (e.g. 
Skilling \& Strong \cite{ski76}). Here, for simplicity, we assume that CRs can freely penetrate the 
clouds if their kinetic energy is higher than a threshold energy $E_{\rm min}$, which is another free 
parameter of the model. We further consider the differential rate of primary CRs that penetrate the 
nonthermal X-ray production region to be a power law in kinetic energy above $E_{\rm min}$: 
\begin{equation} 
{dQ_i \over dt} (E) = C_i E^{-s}~~{\rm for~}E > E_{\rm min}~.
\label{eq7}
\end{equation} 

The model finally has four free parameters that can be studied from spectral fitting of X-ray data 
(see Sect.~5): $\Lambda$, $E_{\rm min}$, the power-law spectral index $s$, and the metallicity of the 
X-ray emitting cloud, $Z$. The X-ray spectral analysis also provides the CR spectrum normalization 
$C_i$, which allows one to estimate the power injected by the primary LECRs into the nonthermal X-ray 
production region: 
\begin{equation} 
{dW_i \over dt} = \int_{E_{\rm min}}^{E_{\rm max}} E' {dQ_i \over dt} (E') dE'~.
\label{eq8}
\end{equation} 
In the following, the integration in the above equation is limited to $E_{\rm max}=1$~GeV. Due to CR 
escape, the power continuously deposited by the fast particles inside the cloud should generally be lower 
than $dW_i/dt$.

\end{appendix}

\begin{appendix}

\section{X-rays from accelerated electron interactions}

In the framework of the adopted steady-state, slab interaction model, the differential X-ray 
production rate from collisions of accelerated electrons with the cloud constituents can be written as
\begin{eqnarray}
{dQ_X \over dt}(E_X) & = & n_{\rm H} \sum_j a_j \int_0^{\infty}  
{d\sigma_{ej} \over dE_X}(E_X,E) v_e(E) \nonumber \\ 
& & \times~ [ N_{e,p}(E) + N_{e,s}(E) ] dE ~,
\label{eq9}
\end{eqnarray}
where $a_j$ is the abundance of element $j$ relative to H in the X-ray emitting cloud,  
$(d\sigma_{ej}/dE_X)$ is the differential X-ray production cross section for 
electron interaction with atoms $j$, and $N_{e,p}$ and $N_{e,s}$ are the 
differential equilibrium numbers of primary and secondary LECR electrons in the ambient 
medium, respectively.  

\subsection{Secondary electron production}

Primary LECR electrons injected into an interstellar molecular cloud produce secondary 
electrons mainly from ionization of ambient H$_2$ molecules and He atoms. The corresponding 
differential production rate of knock-on electrons is given by 
\begin{eqnarray} 
{dQ_{e,s} \over dt} (E_s) & = & n_{\rm H} \int_{2E_s}^{\infty} \bigg[ 
0.5 {d\sigma_{\rm H_2} \over dE_s}(E_s,E_p) + a_{\rm He} {d\sigma_{\rm He} \over
dE_s}(E_s,E_p) \bigg] \nonumber \\
& & \times~ v_e(E_p) N_{e,p}(E_p) dE_p~,
\label{eq10}
\end{eqnarray} 
where $(d\sigma_{\rm H_2}/dE_s)$ and $(d\sigma_{\rm He}/dE_s)$ are the H$_2$ and He 
differential ionization cross sections for the production of a secondary
electron of energy $E_s$ by impact of a primary electron of energy
$E_p$. The lower limit of the integral is $2E_s$, because the primary electron 
is by convention the faster of the two electrons emerging from the collision. 
The maximum possible energy transfer is therefore $E_s=0.5(E_p-B_j) \simeq 
0.5E_p$, where $B_{\rm H_2}=15.43$~eV and $B_{\rm He}=24.59$~eV are the electron 
binding energies of H$_2$ and He, respectively. This convention is
consistent with the definition of the stopping powers used throughout this paper 
(see Eq.~(\ref{eq3})), which also pertain to the outgoing electron of higher energy. 

The differential ionization cross sections are calculated from the 
relativistic binary encounter dipole (RBED) theory (Kim et al. 
\cite{kim94,kim00b}). This successful model combines the binary-encounter 
theory for hard collisions with the dipole interaction of the Bethe theory for 
fast incident electrons. For the differential oscillator strengths, we use 
the analytic fits provided by Kim et al. (\cite{kim94}) for H$_2$ and Kim et al. 
(\cite{kim00a}) for He. For the average orbital kinetic energy of the 
target electrons, we take $U_{\rm H_2}=15.98$~eV and $U_{\rm He}=39.51$~eV.

By inserting Eq.~(\ref{eq1}) into Eq.~(\ref{eq10}) and using for the electron energy
loss rate the expression given in Eq.~(\ref{eq3}), we see that the secondary electron 
production rate does not depend on the absolute density of H atoms in the ambient 
medium ($n_{\rm H}$). This comment also applies to the X-ray production rate, which 
only depends on the relative abundances $a_j$ (see Eq.~(\ref{eq9})). This important 
property of the adopted steady-state, slab model will allow us to estimate unambiguously
the cosmic-ray power $dW_i/dt$ (Eq.~\ref{eq8}) from the measured X-ray flux. 

   \begin{figure}
   \centering
   \includegraphics[width=7.5cm]{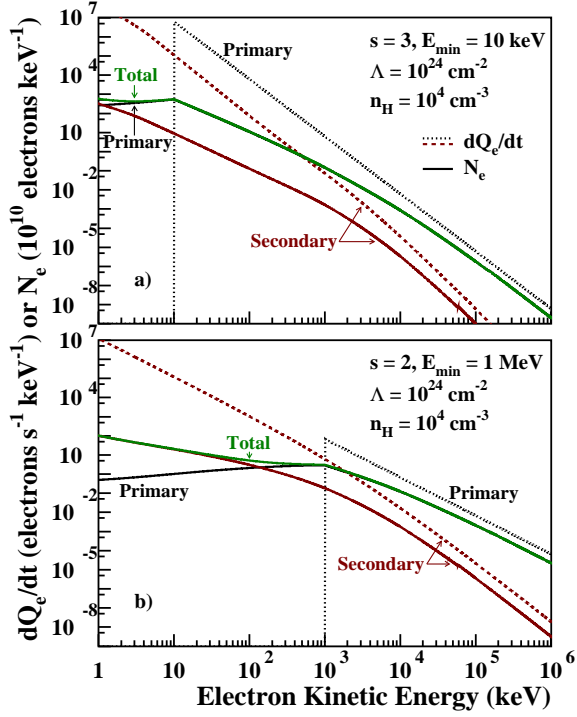}
      \caption{Calculated differential equilibrium electron numbers ($N_{e}$; 
      solid lines) for two differential injection rates of primary electrons 
      ($dQ_{e,p}/dt$; dotted lines): {\bf a)} $s$=3, $E_{\rm min}$=10 keV; {\bf b)} $s$=2,
      $E_{\rm min}$=1 MeV. Also shown are the differential production rates of 
      secondary, knock-on electrons ($dQ_{e,s}/dt$; dashed lines). The H density 
      in the nonthermal X-ray production region, which intervenes in the calculation 
      of $N_{e,p}$ and $N_{e,s}$, is $n_{\rm H}=10^4$~cm$^{-3}$  and 
      the path length of the primary electrons in this region is $\Lambda=10^{24}$~cm$^{-2}$. 
      The calculations are normalized to a total power of 1~erg~s$^{-1}$ injected by 
      the primary LECR electrons in the X-ray production region. 
              }
         \label{Fig2}
   \end{figure}

Calculated differential production rates of primary and knock-on electrons are shown in
Fig.~\ref{Fig2}. Also shown is the corresponding steady-state differential number 
of secondary electrons in the ambient medium, $N_{e,s}$. We calculated the latter from 
Eqs.~(\ref{eq1}) and (\ref{eq2}), assuming the characteristic escape path length of the 
secondary particles to be $\Lambda$/2. Although this assumption is uncertain, it has 
no significant effect on the total X-ray production. 

We see in Fig.~\ref{Fig2} that the effect of H$_2$ and He ionization on the electron 
energy distribution is to redistribute the total kinetic energy of the injected 
particles to a larger number of lower-energy electrons. Thus, for hard enough primary 
electron spectrum (i.e. low $s$ and high $E_{\rm min}$, see Fig.~\ref{Fig2}b), secondary 
electrons of energies $E_s \gsim 10$~keV could potentially make a significant contribution to 
the total nonthermal X-ray emission. On the other hand, one can easily check that 
the successive production of knock-on electrons by the secondary electrons themselves can be 
safely neglected for the X-ray emission.

\subsection{X-ray continuum emission}

The X-ray continuum emission is due to the bremsstrahlung of both primary and 
secondary electrons. We take electron bremsstrahlung into account only in ambient 
H and He and calculate the differential cross sections from the work of 
Strong et al. (\cite{str00}, appendix A), which is largely based on Koch \& 
Motz (\cite{koc59}). We use the scattering functions from Blumenthal \& Gould 
(\cite{blu70}) to take into account the arbitrary screening of the H and He 
nuclei by the bound electrons. 

\subsection{X-ray line emission}

   \begin{figure}
   \centering
   \includegraphics[width=7.5cm]{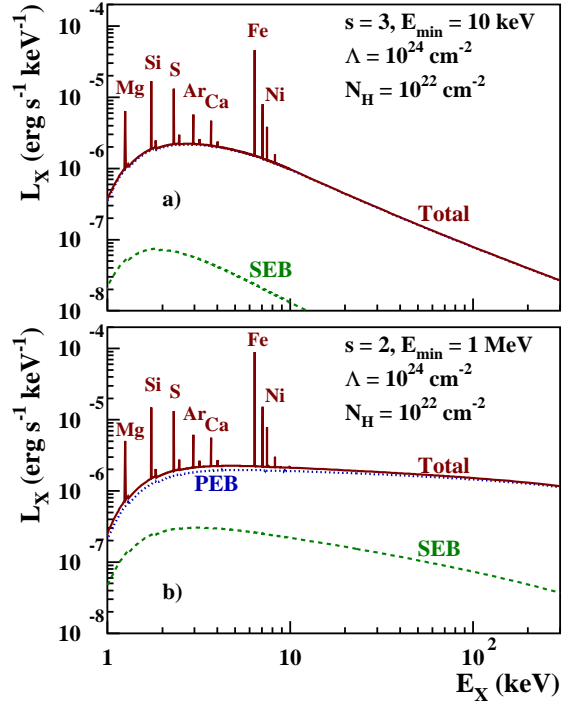}
      \caption{Calculated X-ray emission produced by LECR electrons with the source spectra 
      shown in Fig.~\ref{Fig2} interacting in a gas cloud of solar metallicity. 
      PEB: primary electron bremsstrahlung; SEB: secondary electron bremsstrahlung. 
      Photoelectric absorption is taken into account with a H column density of 
      10$^{22}$~cm$^{-2}$.
              }
         \label{Fig3}
   \end{figure}

The X-ray line emission results from the filling of inner-shell vacancies 
produced by fast electrons in ambient atoms. We consider the K$\alpha$ 
and K$\beta$ lines (2$p$$\rightarrow$1$s$ and 3$p$$\rightarrow$1$s$ transitions 
in the Siegbahn notation) from ambient C, N, O, Ne, Mg, Si, S, Ar, Ca, Fe, and 
Ni. The corresponding cross sections can be written as
\begin{equation}
\sigma_{ej}^{Kk}(E) \equiv {d\sigma_{ej}^{Kk} \over dE_X}(E_X,E) = \delta(E_X-E_{Kk}) 
\sigma_{ej}^I(E) \omega_j^{Kk}~,
\label{eq11}
\end{equation}
where $E_{Kk}$ is the energy of line K$k$ (\kalpha or K$\beta$),
$\delta(E_X-E_{Kk})$ is Dirac's delta function, $\sigma_{ej}^I(E)$ the total
cross section for the K-shell ionization of atom $j$ by an electron of energy
$E$, and $\omega_j^{Kk}$ the K$k$ fluorescence yield for atom $j$ (Kaastra \& 
Mewe \cite{kaa93}). Note that $\omega_j^{K\beta}$=0 for element $j$ with atomic 
number $\leq 12$ (i.e. Mg), since these atoms do not have 3$p$ electrons in 
their ground level. 

For the K-shell ionization cross sections, we adopted the semi-empirical 
formula of Quarles (\cite{qua76}), which agrees well with the RBED cross 
sections for Ni and lighter elements (see Kim et al. \cite{kim00b}) and is
simpler to use. We checked that the Quarles's formula correctly reproduces
the data compiled in Long et al. (\cite{lon90}), in particular at relativistic
energies. 

The width of the X-ray lines produced by electron impact can be estimated from 
the sum of the natural widths of the atomic levels involved in the transition. 
Indeed, broadening effects caused by multiple simultaneous ionizations can 
be safely neglected for LECR electrons. Thus, the K$\alpha_1$ and K$\alpha_2$ 
components of the Fe \kalpha line have experimental full widths at 
half-maximum (FWHM) of only 2.5 and 3.2~eV, respectively (Salem \& Lee 
\cite{sal76}). However, the energy separation of the two fine-structure 
components is 13~eV, which is much less than the energy resolution at 
6.4~keV of the X-ray cameras aboard \textit{XMM-Newton} and \textit{Chandra}, 
but larger than the expected resolution of the \textit{ASTRO-H} X-ray Calorimeter 
Spectrometer (7~eV FWHM; Takahashi et al. \cite{tak10}). 
Here, we neglect the fine-structure splitting of the K lines and for simplicity adopt 
the same width for all the lines: $\Delta E_X = 10$~eV. 

Figure~\ref{Fig3} shows calculated nonthermal X-ray spectra ($L_X=E_X \times dQ_X/dt$) 
produced by LECR electrons injected with the differential rates shown in Fig.~\ref{Fig2} 
into a cloud of solar metallicity. We took the photoelectric absorption of 
X-rays into account using a H column density $N_{\rm H}=10^{22}$~cm$^{-2}$ and the cross sections of 
Morrison \& McCammon (\cite{mor83}). We see in Fig.~\ref{Fig3} that the most prominent 
line is that of Fe at 6.40~keV. This is because this element has the highest product of 
\kalpha fluorescence yield ($\omega_{Fe}^{K\alpha}$=0.3039, Kaastra \& Mewe \cite{kaa93}) 
and cosmic abundance. The EW of the Fe \kalpha line is equal to
293 and 394~eV in the spectra shown in panels~a and b, respectively. The second 
strongest line in these spectra is the Si \kalpha line at 1.74~keV; its EW is equal to 
80 and 90~eV in panels~a and b, respectively. We also see in this figure that 
(i) the shape of the continuum emission reflects the hardness of the 
primary electron injection spectrum, and (ii) the total X-ray emission is dominated by the 
contribution of the primary electrons. The emission from the secondary electrons is 
negligible in panel~a and accounts for 10--20~\% of the total emission below 10~keV in 
panel~b. 

\end{appendix}

\begin{appendix}

\section{X-rays from accelerated ion interactions}

The differential X-ray production rate from accelerated ion interactions can be written 
with a slight modification of Eq.~(\ref{eq9}), as follows:
\begin{eqnarray}
{dQ_X \over dt}(E_X) & = & n_{\rm H} \sum_j a_j \int_0^{\infty}  
\bigg[ \sum_i {d\sigma_{ij} \over dE_X} (E_X,E) v_i(E) N_i(E) \nonumber \\ 
& & + {d\sigma_{ej} \over dE_X}(E_X,E) v_e(E) N_{e,s}(E) \bigg] dE ~,
\label{eq13}
\end{eqnarray}
where the index $i$ runs over the constituents of the nonthermal ion population. The 
first term in the integral represents the X-ray production by the primary 
LECR ions and the second term the contribution of the secondary electrons. As a 
starting point, we assume in the present work that the LECR ion population is 
mainly composed of protons and $\alpha$ particles and that the contributions of 
accelerated metals to the total X-ray emission can be neglected. We therefore
do not consider the broad X-ray line emission that can arise from atomic transitions 
in fast C and heavier species following electron captures and excitations (Tatischeff 
et al. \cite{tat98}), except in Sect.~7.2. However, for typical compositions of accelerated cosmic particles, the fast 
metals significantly contribute neither to the production of the X-ray lines 
from the ambient atoms nor to the bremsstrahlung continuum radiation (see Tatischeff et 
al. \cite{tat98}). We further assume that the accelerated protons and $\alpha$ 
particles are in solar proportion, that is, $C_\alpha / C_p = a_{\rm He}$ 
(see Eq.~(\ref{eq7})). 

In the calculations of the equilibrium spectra ($N_p$ and $N_\alpha$), we neglect the 
nuclear destruction and catastrophic energy loss (e.g. interaction involving pion
production) of the fast ions in the cloud. Indeed these processes are not important 
in comparison with the ionization losses below $\sim$300~MeV~nucleon$^{-1}$ kinetic 
energy (see, e.g., Schlickeiser \cite{sch02}) and most of the X-ray emission below 
10~keV, which is the prime focus of the present work, is produced by ions in this 
low energy range (see Fig.~\ref{Fig8}b).

\subsection{Secondary electron production}

   \begin{figure}
   \centering
   \includegraphics[width=7.5cm]{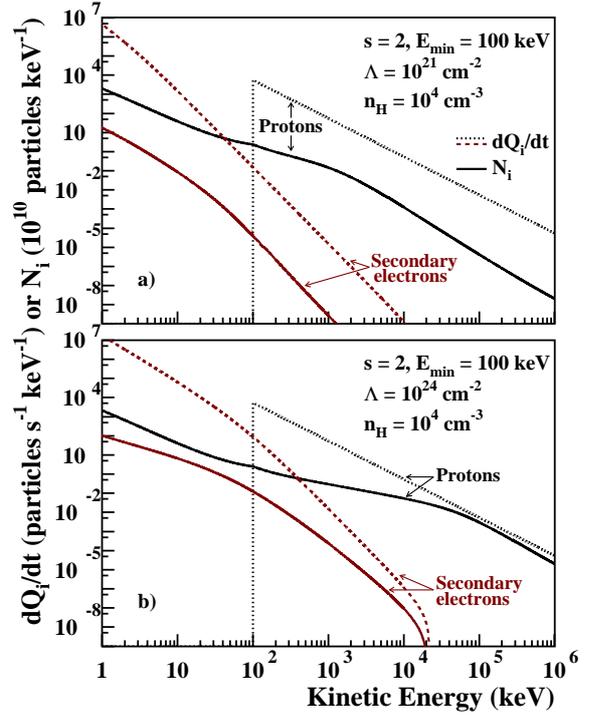}
      \caption{Calculated differential equilibrium numbers of fast particles 
      ($N_i$; solid lines) for the differential injection rate of primary 
      protons given by $s$=2 and $E_{\rm min}=100$~keV ($dQ_p/dt$; dotted lines). 
      Also shown are the differential production rates of secondary knock-on electrons 
      ($dQ_{e,s}/dt$; dashed lines). {\bf a)}: $\Lambda=10^{21}$~cm$^{-2}$; 
      {\bf b)}: $\Lambda=10^{24}$~cm$^{-2}$. The H density in the nonthermal 
      X-ray production region is $n_{\rm H}=10^4$~cm$^{-3}$. The calculations 
      are normalized to a total power of 1 erg~s$^{-1}$ injected by 
      the primary LECR protons in this region.       
	      }
         \label{Fig9}
   \end{figure}

We calculate the production of secondary electrons associated to the ionization of 
ambient H$_2$ molecules and He atoms. The corresponding differential ionization 
cross sections are obtained as in Tatischeff et al. (\cite{tat98}) from the work of 
Chu et al. (\cite{chu81}). We neglect the production of secondary electrons and 
positrons that follows the production of charged pions in hadronic collisions. 
In fact, the corresponding electron and positron source functions can dominate the 
one of knock-on electrons only at energies $> 10$~MeV (Schlickeiser \cite{sch02}), 
and these high-energy leptons are not important for the production of X-rays $< 10$~keV 
(see Fig.~\ref{Fig8}a).

Differential production rates of knock-on electrons are shown in Fig.~\ref{Fig9}, 
together with the corresponding equilibrium spectra of primary protons and secondary 
electrons. This figure illustrates the effects of changing the CR escape path length 
from $\Lambda=10^{21}$~cm$^{-2}$ (panel a) to $10^{24}$~cm$^{-2}$ (panel b). In the first 
case, protons of energies up to 1.4~MeV are stopped in the cloud, whereas in the second 
case the transition energy between proton stopping and escape is at 71~MeV. We see that 
above this transition energy the equilibrium spectrum has a similar slope than the source 
spectrum, whereas at lower energies the equilibrium proton distribution is harder due to 
the ionization losses. We can anticipate that the total X-ray production rate will be much 
higher for the case $\Lambda=10^{24}$~cm$^{-2}$, as a result of the higher proton number at 
equilibrium above a few MeV. 

\subsection{X-ray continuum emission}

The X-ray continuum emission is due to inverse bremsstrahlung from the fast ions (the 
radiation of a single photon in the collision of a high-speed ion with an electron 
effectively at rest) and classical bremsstrahlung from the secondary knock-on electrons. 
In the nonrelativistic domain, the bremsstrahlung produced by a proton of kinetic energy 
$E$ in a collision with a H atom at rest has the same cross section as that of an electron 
of kinetic energy $(m_e/m_p)E$ in a collision with a stationary proton ($m_e$ and $m_p$ 
are the electron and proton masses, respectively). We calculate this cross section as in 
Sect.~3.2, but without taking the screening of the H nucleus by the bound electron into 
account. The cross section for interaction of a proton with a H atom is then multiplied 
by $(1+2a_{\rm He})$ to take the ambient He into account. For $\alpha$ particles, we replace 
the proton energy $E$ by the energy per nucleon of the projectile and multiply the proton 
cross section by 4 to account for the nuclear charge dependence of the bremsstrahlung cross 
section. 

In the relativistic case, the cross section for proton inverse bremsstrahlung is different 
from the one for classical electron bremsstrahlung, owing to the appearance of angular and energy 
abberations in the transformation between the two rest frames of the interacting particles 
(Haug~\cite{hau03}). We checked that these effects can be neglected in good approximation 
in the present work. 

\begin{figure}
   \centering
   \includegraphics[width=7.5cm]{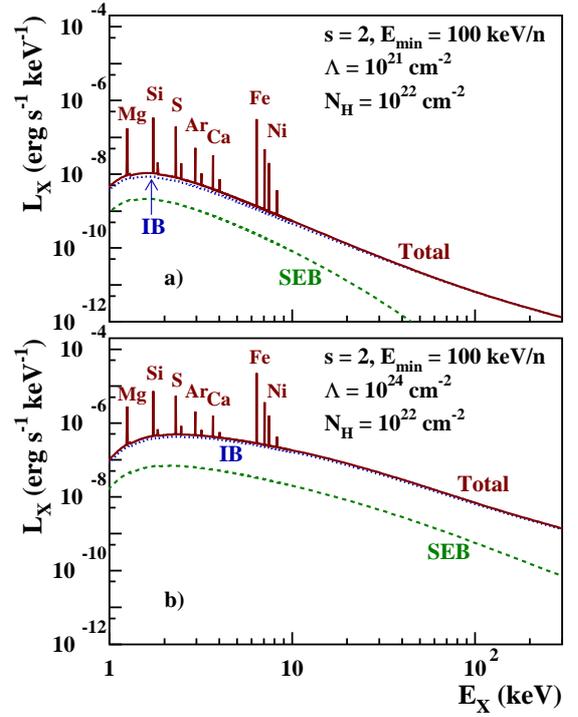}
      \caption{Calculated X-ray emission produced by LECR protons and $\alpha$-particles 
      interacting in a gas cloud of solar metallicity, for the differential injection 
      rate of primary protons shown in Fig.~9. The contribution of accelerated $\alpha$-particles 
      is included as explained in the text, assuming in particular the solar abundance 
      $C_\alpha / C_p = 0.0964$. {\bf a)}: $\Lambda=10^{21}$~cm$^{-2}$; 
      {\bf b)}: $\Lambda=10^{24}$~cm$^{-2}$.  IB: inverse bremsstrahlung; SEB: secondary 
      electron bremsstrahlung. Photoelectric absorption is taken into account with a H 
      column density of 10$^{22}$~cm$^{-2}$.
              }
         \label{Fig10}
   \end{figure}

In Fig.~\ref{Fig10} we show two X-ray spectra corresponding to the particle equilibrium 
spectra presented in Fig.~\ref{Fig9}. We see that the continuum emission is dominated by 
inverse bremsstrahlung, which is a general rule independent of the model parameters (see Tatischeff 
et al. \cite{tat98}). We also see that, as expected, the X-ray production rate is much higher 
for $\Lambda=10^{24}$~cm$^{-2}$ than for $\Lambda=10^{21}$~cm$^{-2}$, the difference being a 
factor of 22, 337 and 1054 at 1, 10, and 100 keV, respectively.

\subsection{X-ray line emission}

For producing X-ray lines from the ambient atoms, we take both the contribution from secondary 
electrons (see Eq.~(\ref{eq11})) and that from primary ions into account. The cross sections for 
K-shell ionization by proton and $\alpha$-particle impacts are extracted from the data library 
implemented by Pia et al. (\cite{pia09}) in the Geant4 toolkit for the simulation of particle 
induced X-ray emission (PIXE). We use the cross sections calculated in the ECPSSR theory with 
high-velocity corrections (Lapicki et al. \cite{lap08}). These cross sections are more 
accurate for mildly relativistic projectiles than those previously employed by Tatischeff et al 
(\cite{tat98}). 

Proton and $\alpha$-particle collisions with target atoms do not lead to significant line broadening 
effects caused by multiple simultaneous ionizations. We thus adopt as before a width of 10~eV for all 
the lines (see Sect.~3.3). We note, however, that the X-ray lines produced by collisions of ions heavier 
than $^4$He can be shifted by several tens of eV, significantly broadened and split up into several 
components (Garcia et al. \cite{gar73}). For example, the Fe \kalpha line produced by impacts of O ions 
of 1.9~MeV~nucleon$^{-1}$ is blueshifted by $\sim$50~eV in comparison with the one produced by 5-MeV proton 
impacts, and has a FWHM of$\sim$100~eV (see Garcia et al. \cite{gar73}, figure 3.55). 

The most intense line produced by LECR protons and $\alpha$-particles is also the neutral Fe 
K$\alpha$ line at 6.40~keV (Fig.~\ref{Fig10}). This line has an EW of 2.31
and 0.80~keV in the spectra shown in Figs.~\ref{Fig10}a and b, respectively. The second 
strongest line in these spectra is the Si \kalpha line at 1.74~keV; its EW is equal to 
309~eV in panel~a and 152~eV in panel~b.

\end{appendix}

\end{document}